\documentclass[preprint,amsmath,amssymb,aps,showkeys,showpacs]{revtex4}
\usepackage[colorlinks=true, pdfstartview=FitV, linkcolor=blue, citecolor=red, urlcolor=magenta]{hyperref}
\usepackage{graphicx}
\usepackage{latexsym}
\usepackage{amsmath}
\usepackage{amsfonts}
\usepackage{amssymb}
\usepackage{verbatim}
\usepackage{dcolumn}
\usepackage{amssymb}
\usepackage{bm}
\usepackage{float}
\usepackage{subfigure}
\usepackage[english]{babel}
\newcommand{\be}{\begin{equation}}
\newcommand{\ee}{\end{equation}}
\newcommand{\bea}{\begin{eqnarray}}
\newcommand{\eea}{\end{eqnarray}}

\begin{document}

\newcommand{\NITK}{
\affiliation{Department of Physics, National Institute of Technology Karnataka, Surathkal  575 025, India}
}

\title{Critical Behaviour and Microscopic Structure of Charged AdS Black Hole with a Global Monopole in Extended and Alternate Phase Spaces}

\author{Naveena Kumara A.}
\email{naviphysics@gmail.com}
\NITK
\author{Ahmed Rizwan C.L.}
\email{ahmedrizwancl@gmail.com}
\NITK
\author{Deepak Vaid}
\email{dvaid79@gmail.com}
\NITK
\author{Ajith K.M.}
\email{ajithkm@gmail.com}
\NITK


\begin{abstract}
A detailed discussion on phase transition and microscopic structure of charged AdS black hole with a global monopole is presented in both extended and alternate phase spaces. In the analysis of critical behaviour, the classical van der Waals analogy is drawn from isotherms which is followed by Gibbs free energy study and coexistence curves. In both spaces, the symmetry breaking parameter $\eta$ acts as a hindrance for critical behaviour. The crux of van der Waals like behaviour is investigated by looking at the microscopic structure of the black hole via thermodynamic Ruppeiner geometry. The Ruppeiner invariant scalar behaves differently in extended and alternate spaces. The monopole parameter influences the microscopic structure of the black hole, which in turn, affects the critical behaviour. The effect is significant at the maximal strength of the monopole parameter.

\end{abstract}

%\includegraphics{alt_coext}




\maketitle

\section{Introduction}
\label{intro}
Black hole physics has changed from mere theoretical importance to experimental aspects due to the observational advances like gravitational waves and black hole imaging. However, the theoretical developments are far ahead of experiments due to its importance in several directions like quantum mechanics, quantum gravity and string theory. Black hole thermodynamics began with a quest for incorporating quantum mechanical nature to a black hole, which had purely classical origin in general relativity. Since the pioneering work in this regard by Hawking and Bekenstein, black hole thermodynamics remains as an exciting topic in contemporary research.

The first step taken in establishing such a domain by Hawking and Bekenstein was by introducing a temperature and entropy for a black hole which are related to surface gravity $\kappa$ and area of the black hole respectively 
\citep{Hawking:1974sw, Bekenstein1972}. The four laws of black hole thermodynamics parallel to classical thermodynamics were soon proposed by taking the mass of the black hole as internal energy \citep{Bekenstein1973, Bekenstein1974, Bardeen1973}. The importance of AdS black holes in black hole thermodynamics was realised in the early stage of these developments from the result that the thermodynamically stable black holes exist only in AdS space. This is in contrast to the Minkowskian case, where the black hole has a negative specific heat and disappears by radiating  Hawking radiation. This happens because the boundary of AdS space acts like walls of a thermal cavity. Inside this closed box-like space, below a certain temperature only radiation can exist. But, above that temperature, the radiation becomes unstable and hence collapses, resulting in the formation of black holes. The black holes thus formed, exists in two forms, larger ones with positive specific heat which are locally stable, and smaller ones with negative specific heat which are unstable. The phase transition between these black holes and radiation at the transition temperature is termed as Hawking Page transition \citep{Hawking1983, Page2005}. In high energy physics, the topic AdS space got the attention of a larger audience after the proposal of AdS-CFT correspondence by Maldacena \cite{Maldacena1999}. This gauge-gravity duality relates gravity theory in an AdS space to the conformal field theories at the boundary of that space. When Hawking Page transition is seen in the light of AdS-CFT language it appears as confinement/deconfinement phase transition in quantum chromodynamics \citep{Witten:1998zw}. 

The first progress beyond Hawking Page phase transition happened after the identification of a rich phase structure isomorphic to van der Waals liquid-gas system in RN-AdS black hole \citep{Chamblin1999, Chamblin:1999hg} and in Kerr RN-AdS black hole \citep{Caldarelli:1999xj}. Till then, in all black hole thermodynamic studies, pressure and volume, the crucial thermodynamic variables were absent. The introduction of pressure to this field is done through cosmological constant $\Lambda$, which also has other fundamental implications like the consistency of Smarr relation with first law \citep{Kastor:2009wy}. In this approach, the conjugate quantity of cosmological constant is taken as the thermodynamic volume. At the same time, the expansion of phase space altered the first law of thermodynamics with a $PdV$ term, giving a new interpretation to the mass of the black hole as enthalpy \citep{Dolan2011}. The new perspective on mass and cosmological constant in black hole thermodynamics resulted in phenomenal consequences, enabling one to establish newer analogies with the well-known phenomena in classical thermodynamics. Understanding AdS black hole as a replica of van der Waals system by Kubiznak \emph{et al.} was a milestone in this regard \citep{Kubiznak2012, Kubiznak2017}. In their work, a detailed account of the criticality of the black hole is presented with $P-v$ isotherm, Gibbs free energy plot, coexistence curve and critical exponents, which are all shown to be similar to van der Waals case. Since then, similar studies were conducted on various AdS black holes on different contexts but with universal features \citep{Gunasekaran2012, BelhajChabab2012, Hendi2013, SChen2013, SPALLUCCI2013, Zhao2013}. Other analogies to classical thermodynamics like Joule Thomson expansion \citep{Okcu2017}, holographic heat engines \citep{Johnson:2014yja}, Clausius-Clapeyron Equation \citep{Zhao:2014fea} and reentrant phase transitions \citep{Altamiranokubi2013} were also made.

Recently, there emerged a novel idea of identifying the square of the charge of the black hole ($Q^2$) with fluid pressure, which differs from earlier extended phase space studies where the cosmological constant is a variable \citep{Dehyadegari2017}. The origin of this alternate phase space traces back to the work of Chamblin \emph{et al.}, much before the introduction of thermodynamics in extended space \citep{Chamblin1999}. Initially, the black hole charge ($Q$) was thought to be reasonable to tune rather than cosmological constant, which has a fixed value in general relativity. For a charged black hole, it is more natural to think of charge as a thermodynamic variable. In these approaches, energy differential corresponding to charge is taken as $\Phi dQ$ with $\Phi =Q/r_+$, which induces the thermodynamics and phase structure, somewhat similar to extended phase space. While identifying the fluid pressure with the charge of the black hole $Q$,  the corresponding conjugate quantity, the electric potential $\Phi$ is taken as the candidate for the volume of the fluid. However, the thermodynamics presented in this new space had some drawbacks related to the response function $dQ/d\Phi$, which lacks the proper physical meaning.  The variables $Q$ and $\Phi$ are not mathematically independent quantities since $\Phi =Q/r_+$. Moreover, the instability associated with the multivaluedness of Gibbs free energy $G$ is not removed in the corresponding $Q-\Phi$ isothermal diagram after the Maxwell construction. This is because both regions with negative and positive slopes exist in spite of Maxwell construction. Since the study of critical behaviour requires the identification of stable and unstable regimes, this phase space is not a useful one. Such suspicious results are overcome in the alternate phase space with $Q^2-\Psi$ isotherms. If we take an arbitrary function of $Q$ then the proper identification of independent conjugate variable is not possible. In the alternate phase space, one can establish the similarities between the thermodynamic behaviour of AdS black holes and van der Waals fluid. In this new approach, one can attribute the phase transition of the black hole to its charge. Here, the mass of the black hole is taken as a function of the square of the charge instead of the charge itself. Therefore, the conjugate variable for $Q^2$ is the inverse of the specific volume, $\Psi = 1/v$. The first law and Smarr formula are also modified accordingly. The first detailed analysis in this regard on charged AdS black hole showed the connection between thermodynamic geometry and critical behaviour, as it was in extended phase space \citep{Dehyadegari2017}. In the successive studies, the effect of quintessence on RN-AdS black hole \citep{Chabab2018},  critical behaviour of Gauss-Bonnet black hole \citep{Yazdikarimi:2019jux} and that of Lifshitz dilaton black hole \citep{Dayyani:2018mmm} via alternative phase space were reported. The universality class and critical properties for generic AdS black holes in general via alternative phase space were also studied and the results were found to be similar to van der Waals system \citep{Dehyadegari:2018pkb}. The phase structure and critical behaviour of Born-Infeld (BI) black holes in AdS space, with the variation of charge of the system and with fixed cosmological constant (pressure) are also studied \citep{Dehyadegari:2017hvd}. In this investigation, it was found that the system admits a reentrant phase transition. 

An effective way of approaching phase transitions in classical thermodynamics is thermodynamic geometry on phase space. The essential part of this method is the construction of a thermodynamic metric. The first such metric is proposed by Weinhold in which hessian of internal energy was the key entity, which is a function of entropy and other extensive quantity \citep{weinhold1975metric}. Later, another construction of metric was given by Ruppeiner, where, instead of internal energy, entropy is taken as generating function in the definition of hessian \citep{Ruppeiner79, Ruppeiner95}. This new construction appears more appropriate for black hole thermodynamics since the entropy of the black hole is measured on the horizon, whereas the internal energy is at the asymptotic infinity. However, one can show that these different constructions are related to each other conformally with the inverse of temperature as a conformal factor. In the fluctuation theory of equilibrium thermodynamics inverse of Ruppeiner metric gives the second moments of fluctuations. At the critical point, thermodynamic scalar curvature becomes proportional to the correlation volume of $\xi$ of the thermodynamic system. The curvature scalar contains the underlying microscopic structure of the corresponding statistical system, which shows diverging behaviour near the critical point. The curvature scalar vanishes for ideal gas since there is no intermolecular interaction. Several applications of thermodynamic geometry on different black hole spacetimes revealed interesting results and newer possibilities \citep{Ferrara:1997tw, Aman2003, bTzSarkar2006, SHEN2007, TSarkar2008, Ruppeinerb2008, Sahay2010, Lala2012, SheykhiEMd2015, lifR2016, Sahay2017}. The interplay between geometry and microscopic structure in the thermodynamic geometric method is extremely useful in black hole thermodynamics since the exact knowledge of constituent information of black hole is still a debatable issue. We would like to emphasize that in both ordinary thermal system and black hole system, thermodynamic geometry gives only a phenomenological description, not the exact microstructure. In our research, we investigate the properties of microstructure only.

The fact that the black hole is a thermal system gives a strong hint about the existence of the microstructure and corresponding microscopic degrees of freedom. As Boltzmann said, “If it is hot, it must have microstructure”. 
Our motivation for this work is the recent interest among the researchers to explore the microscopic structure of charged AdS black holes using thermodynamic geometry. 
 By introducing a concept called number density\cite{Wei2015}, its found that black holes are composed of effective molecules carrying the microscopic degrees of freedom of its entropy. Another study found that the interaction between the black hole micromolecules can be described by using  Lennard-Jones potential \cite{Miao2017}. This study also reports that the microscopic behaviour of the small black hole exactly matches with that of the ideal anyon gas and that of a large black hole matches with the ideal Bose gas. Later, studying the effect of the scalar curvature of thermodynamic geometry on the phase transition in the binary fluid model, it is observed that the microstructure of charged molecules is like a fermion gas and that of uncharged molecules is like a boson gas \cite{Guo2019}.  In the same work, an analogy with anyon gas with repulsive and attractive interaction is also proposed for the average interaction of black hole molecules. The microstructure point of view is also used to analyse the reentrant phase transitions and it is reported that the small black hole behaves like a bosonic gas, in contrast to their anyon gas like behaviour in the usual phase transition case \cite{Zangeneh2017}. The microscopic interaction of a black hole is different from that of van der Waals fluid. While the latter possess only dominant attractive interaction, the former shows a dominant repulsive interaction in a small parameter range along with the attractive interaction \cite{Wei2019a}. The above properties on the microstructural aspects of the black hole are observed in different spacetime backgrounds and ensembles\cite{Wei2019b, Wei2019, Dehyadegari2017, Chabab2018, Xu2019, Deng2017, Miao2019, Miao2019a, Chen2019, Du2019}. We hope that these phenomenological studies will guide us to identify the correct theory of quantum gravity in future. Keeping this in mind, we try to address the microstructure of a charged AdS black hole with a dented spacetime due to the presence of a global monopole. The presence of a monopole induces a solid angle deficit around the black hole spacetime, even though it's gravitational effects are negligible.

Monopoles are one among the defects like textures, domain walls and cosmic strings, which are formed during the cooling phase of the early universe \citep{Kibble1976, Vilenkin1985}. These topological defects are the consequence of a  non-uniform spontaneous symmetry breaking. Geometrically these defects are the result of the impossibility of the shrinking the vacuum manifold into a single point. Global monopoles are formed during the symmetry breaking of a self coupled triplet scalar field of $SO(3)$ gauge symmetry spontaneously broken into $U(1)$ gauge. The energy density of these global monopoles has a functional dependence on radial distance as $1/r^2$ and exhibit a solid angle deficit of $8\pi ^2 \eta ^2$ (where $\eta$ is the scale of gauge symmetry breaking). The static black hole solution with a global monopole was first obtained by Barriola and Vilenkin \citep{Barriola1989}, the topological structure of which is distinct compared to the Schwarzschild black hole solution.  The physical properties of this black hole solution with monopole are studied extensively \citep{Shi1991, Banerjee1996, Chen2013, Jusufi2017}. The solid angle deficit possessed by this solution is analogous to the conical deficits due to cosmic strings. Recently the thermodynamics of black holes with conical defects has been investigated in the context of accelerating black holes \citep{Appels:prl, Appels:2017xoe}. The effect of global monopole in superconductor/normal metal phase transition was first investigated by Chen et.al. \citep{Chen:2009vz}. Later, in the spacetime of this monopole black hole, the thermodynamics and phase transitions are observed \citep{Deng2018} and Joule Thomson effect was also studied \citep{AhmedRizwan:2019yxk}. In all these studies the global monopole showed its presence by affecting the phenomena under consideration significantly. Motivated by these results, in this paper, we study a charged AdS black hole with a global monopole.

The motivation for this research is to find the possible change in critical behaviour and the microstructure of the charged AdS black hole due to the solid angle deficit induced by the global monopole. If there exists a relation between the solid angle deficit and microstructure, it is interesting because a similar solid angle deficit is exhibited in Skyrme black hole \citep{Flores_Alfonso_2019} and conical defect in accelerating black hole \citep{Appels:2017xoe}. It is reasonable to ask in the context of AdS-CFT correspondence, what happens to the underlying microstructure of the black hole if the surface properties are changed owing to solid angle deficit. 

This article is organized as follows. In section \ref{monopole} we present a brief overview of the charged AdS black hole with a global monopole. This is followed by the study of thermodynamics in extended and alternate spaces in sections \ref{Phase extended} and \ref{Phase alternate} respectively. The microscopic structure of the black hole is investigated via thermodynamic Ruppeiner geometry in both the spaces in section \ref{TD Geometry}. The paper ends with section \ref{last section} where we discuss our observations and results.

\section{The charged AdS black hole with a global monopole}
\label{monopole}

We begin this section by reviewing the details of charged AdS black hole with a global monopole. The Lagrangian density that characterises the simplest model with a global monopole is \citep{Barriola1989},
\begin{equation}
\mathcal{L}_{gm}=\frac{1}{2}\partial _\mu \Phi ^j \partial  ^\mu {\Phi ^*} ^j-\frac{\gamma}{4}\left( \Phi ^j{\Phi ^*}^j-\eta _0 ^2\right)^2,
\end{equation}
where $\Phi ^j$ is self coupled scalar field triplet, $\gamma$ is a self interaction term and $\eta _0$ is the energy scale of gauge symmetry breaking. The field  configuration  for the scalar triplet which describe the monopole is,
\begin{equation}
\Phi^j= \eta _0 h(r)\frac{x^j}{r}
\end{equation}
where $x^j=\{r \sin \theta \cos \phi , r \sin \theta \sin \phi ,r \cos \theta \}$ with $x^j x^j=r^2$. The generic metric ansatz for static spherically symmetric spacetime is,
\begin{equation}
d\tilde{s}^2=-\tilde{f}(\tilde{r})d\tilde{t}^2+\tilde{f}(\tilde{r})^{-1}d\tilde{r}^2+\tilde{r}^2 d\Omega ^2
\label{ansatz}
\end{equation}
where $d\Omega ^2=d\theta ^2 +\sin ^2 \theta d\phi ^2$. The energy-momentum tensor can be obtained from the Lagrangian density, which is given by,
\begin{equation}
T_{\mu\nu}=\frac{2}{\sqrt{-g}}\frac{\partial}{\partial g^{\mu\nu}}\left(\mathcal{L}_{gm}\sqrt{-g}\right)=2\frac{\partial\mathcal{L}_{gm}}{\partial g^{\mu\nu}}-g_{\mu\nu}\mathcal{L}_{gm}.  
\label{stress energy}
\end{equation}
The explicit forms of the components of $T_{\mu \nu}$ are,
\begin{align*}
T_{tt}&=f(r)\left( \frac{\eta _0^2 {h'}^2 f(r)}{2} +\frac{\eta _0 ^2 h^2}{r^2}+\frac{1}{4}\gamma \eta _0 ^4(h^2-1)^2 \right)\\
T_{rr}&=\frac{1}{f(r)} \left(- \frac{\eta _0^2 {h'}^2 f(r)}{2} +\frac{\eta _0 ^2 h^2}{r^2}+\frac{1}{4}\gamma \eta _0 ^4(h^2-1)^2 \right)\\
T_{\theta \theta}&=r^2 \left( \frac{\eta _0^2 {h'}^2 f(r)}{2}+\frac{1}{4}\gamma \eta _0 ^4(h^2-1)^2 \right)\\
T_{\phi \phi}&=r^2 \sin ^2 \theta \left( \frac{\eta _0^2 {h'}^2 f(r)}{2}+\frac{1}{4}\gamma \eta _0 ^4(h^2-1)^2 \right).
\end{align*}

We are using an approximate solution for the charged AdS black hole with a global monopole \citep{Barriola1989}. The solution for the field equation corresponding to scalar action in curved space is approximated with that of flat space. This approximation is done assuming the structure of the monopole is not affected by gravity in a small range. 

We obtained the field equation for the scalar action in the curved spacetime which reads as follows,

\begin{equation}
\tilde{f}'' h{''}+\frac{2}{r}\tilde{f} h'+\tilde{f}'h' - \frac{2}{r^2}h-\lambda \eta _0^2 h(h^2-1)=0.
\end{equation}
In the flat spacetime this reduces to
\begin{equation}
h{''}+\frac{2}{r}h'-\frac{2}{r^2}h-\frac{h(h^2-1)}{\delta ^2}=0
\label{flat}
\end{equation}
where $\delta \approx (\eta _0 \sqrt{\lambda})^{-1}$ is the monopole core size in the flat space. At small distances,  gravity does not substantially change the structure of the monopole for $\eta _0 < m_p$ (where $m_p$ is the Planck mass),  so that the flat space estimation of $\delta$  applies in curved space. The profile of $h(r)$ in flat space can be obtained by the method of hyperbolic functions by setting $x=r/\delta$ in equation (\ref{flat}) as follows \citep{Shi_1991}

\begin{equation}
 h(x)=\sum _{n=0} ^{\infty} c_n \tanh ^{2n+1} \frac{x}{\sqrt{2}}
\end{equation}
where $c_n$ are the expansion coefficients. 

From the profile of $h(x)$ it is apparent that, $h(r)$ linearly increases for $r<(\eta _0 \sqrt{\gamma})^{-1}$ and exponentially approaches to unity when $r>(\eta _0 \sqrt{\gamma})^{-1}$. So, one can approximate $h(r)\approx1$ outside the monopole core \citep{Barriola1989}. This is also a valid assumption in our calculation because the core is inside the black hole and thermodynamics is studied on the horizon. Secondly, outside the core, the potential corresponding to monopole charge varies as $1/r$, which makes $h(r)\approx1$. With this approximation, the components of energy-momentum tensors are reduced to 

\begin{equation}
T^t_t\approx T^r_r\approx \eta ^2/r^2 \quad , \quad T^\theta _\theta \approx T^\phi _\phi \approx 0.
\end{equation}

This is in agreement with the second argument presented above in terms of potential. The energy depends on the square of the field. The useful Einstein equation for the  spherically symmetric static metric is

\begin{eqnarray}
f\left( \frac{1}{r^2}-\frac{1}{r}\frac{f}{f'}\right) -\frac{1}{r^2}=8\pi G T^t_t\\
f\left( \frac{1}{r^2}+\frac{1}{r}\frac{f'}{f}\right) -\frac{1}{r^2}=8\pi G T^r_r.
\end{eqnarray}

The solution of these differential equations is the required form of $\tilde{f}(\tilde{r})$, 
\begin{equation}
\tilde{f}(\tilde{r})=1-8\pi \eta _0^2 -\frac{2\tilde{m}}{\tilde{r}}.
\end{equation}

Incorporating the charge for the black hole in four dimensional AdS space, the function $f(r)$ of the metric takes the form,
\begin{equation}
\tilde{f}(\tilde{r})=1-8\pi \eta _0^2 -\frac{2\tilde{m}}{\tilde{r}}+\frac{\tilde{q}^2}{\tilde{r}^2}+\frac{\tilde{r}^2}{l^2}.
\end{equation}

Where $\tilde{m}$, $\tilde{q}$ and $l$ are the mass parameter, electric charge parameter and AdS radius of the black hole respectively. Under the following coordinate transformations,
\begin{equation}
\tilde{t}=(1-8\pi \eta _0^2)^{-1/2}t~~,~~\tilde{r}=(1-8\pi \eta _0^2)^{1/2}r,
\end{equation}
and introducing new parameters
\begin{equation}
m=(1-8\pi \eta _0^2)^{-3/2}\tilde{m}~,~q=(1-8\pi \eta _0^2)^{-1}\tilde{q}~,~\eta ^2=8\pi \eta _0^2,
\end{equation}
we have the line element
\begin{equation}
ds^2=-f(r)dt^2+f(r)^{-1}dr^2+ar^2 d\Omega ^2,
\end{equation}
with
\begin{equation}
f(r)=1 -\frac{2m}{r}+\frac{q^2}{r^2}+\frac{r^2}{l^2}, \quad \textrm{and} \quad a=1-\eta ^2.
\end{equation}
The spacetime described by the above metric exhibits a solid angle deficit i.e. the area of the sphere is $4\pi (1-\eta ^2) r^2$ instead of $4\pi r^2$. The electric charge $(Q)$ and the Arnowitt-Deser-Misner (ADM) mass $(M)$ can be expressed as 
\begin{equation}
Q=aq~~,~~M=a m.
\end{equation} 
At the event horizon $r=r_+$ the function $f(r)$ vanishes. This condition can be used to determine the mass parameter. The ADM mass now has the following form,
\begin{equation}
M=\frac{ar_+}{2}+\frac{Q^2}{2ar_+}+\frac{ar_+^3}{2l^2}.
\label{eq9}
\end{equation}

\section{Thermodynamics in Extended Phase Space}
\label{Phase extended}
In the extended phase space, the first law of thermodynamics and Smarr relation  reads as follows
\begin{equation}\label{eq10}
dM=TdS+\Phi dQ+VdP~~,~~M=2(TS-PV)+\Phi Q.
\end{equation} 
The existence of global monopole does not affect the form of first law of thermodynamics and the Smarr relation. This can be easily verified by using Euler's homogeneous function theorem. However, global monopole will change the thermodynamics through it's explicit appearance in the thermodynamical quantities \citep{Deng2018}. The crucial thermodynamic variable to begin with is the  entropy $S$ of the black hole, which is related to the area $A_{bh}$ of the event horizon,
\begin{equation}\label{eq11}
S=\frac{A_{bh}}{4}=\pi ar_+^2.
\end{equation}
The soul of extended phase space lies in the identification  of cosmological constant $(\Lambda)$ with the thermodynamic variable pressure $(P)$, and association of its conjugate quantity with the thermodynamic volume $(V)$,
\begin{equation}
P=-\frac{\Lambda}{8\pi}=\frac{3}{8\pi l^2}~~,~~V=\frac{4}{3}\pi ar_+^3.
\label{eqP}
\end{equation}
Using equation (\ref{eq9}) in first law the Hawking temperature $(T)$ for the black hole is an immediate result,
\begin{equation}
T=\left( \frac{\partial M}{\partial S}\right) _{P,Q}=\frac{1}{4\pi r_+}\left(1+\frac{3r_+^2}{l^2}-\frac{Q^2}{a^2r_+^2}\right).
\label{eqT}
\end{equation}
\begin{figure}[H]
    \subfigure[]
    {
        \includegraphics[width=0.5\textwidth]{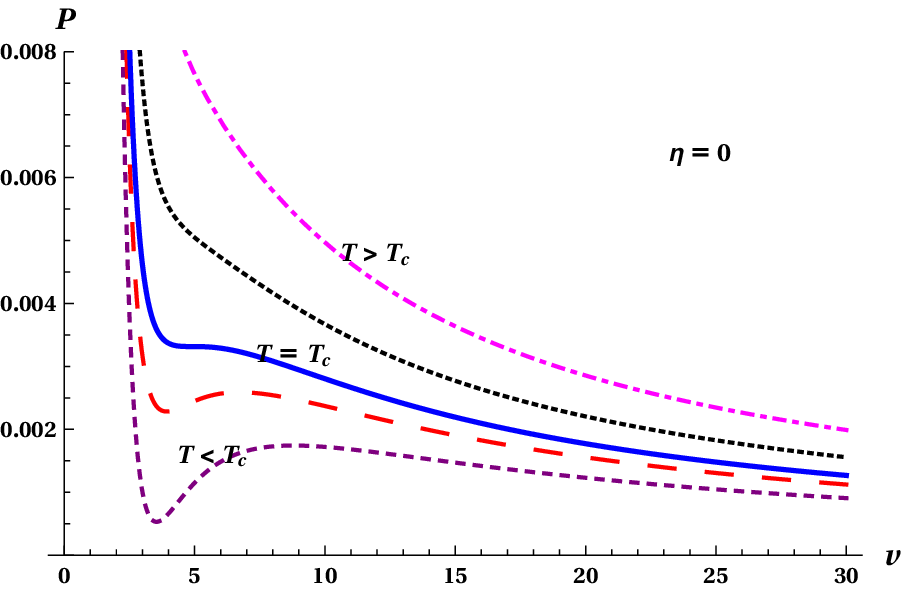}
        \label{PV1}
    }
    \subfigure[]
    {
        \includegraphics[width=0.5\textwidth]{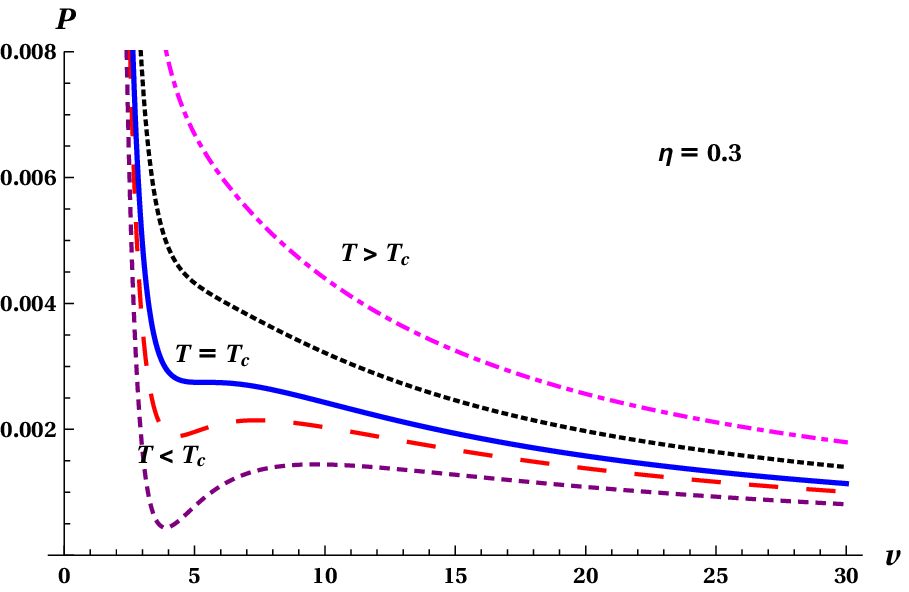}
        \label{PV2}
    }
       
        \subfigure[]
    {
        \includegraphics[width=0.5\textwidth]{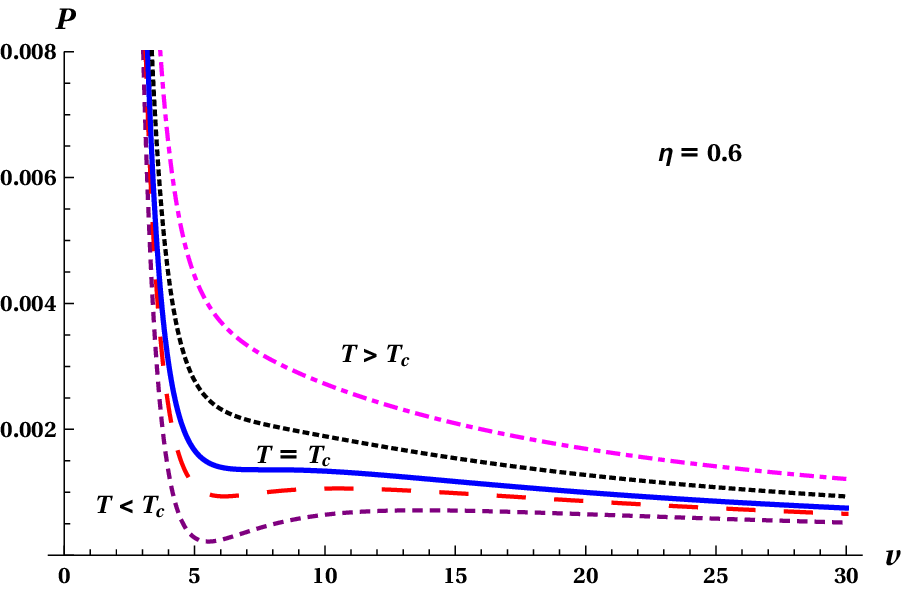}
        \label{PV3}
    }
    \subfigure[]
    {
        \includegraphics[width=0.5\textwidth]{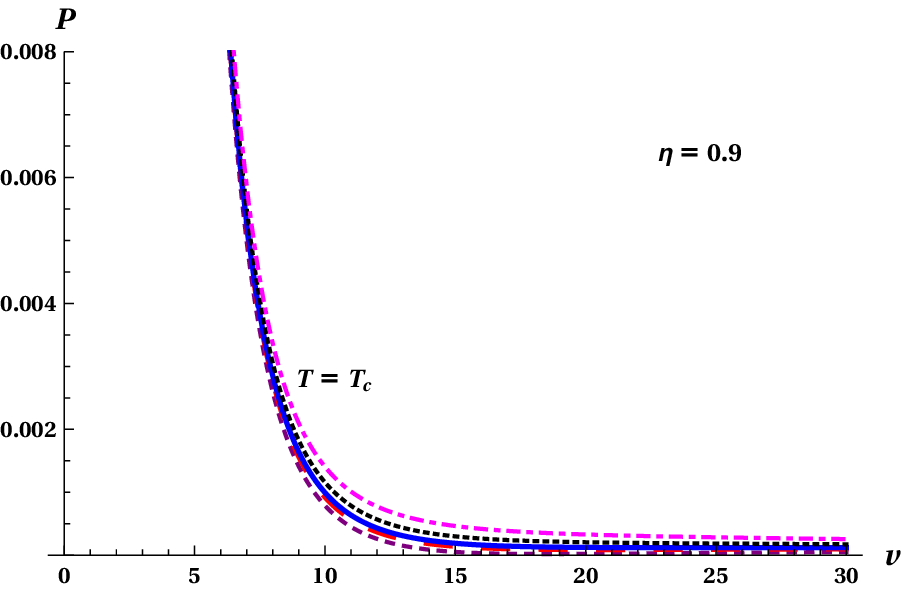}
        \label{PV4}
    }
\\
\caption{$P-v$ isotherms in extended phase space. The critical behaviour is seen below a critical temperature $T_C$. This behaviour reduces with increase in $\eta$. In all these plots temperature is in decreasing order from top to bottom. (We have set $Q=1$ in every plot).}\label{PVs}
\end{figure}

So far all the thermodynamic variables are modified in the presence of global monopole, which has a message within it about its role in influencing the thermodynamics of the black hole. Since a charged black hole is analogous to van der Waals fluid with similar critical behaviour, the tuning of $\eta$ will enhance or suppress the phase transition \citep{Deng2018}.  

Plugging equation (\ref{eqP}) into (\ref{eqT}) eliminates the cosmological constant and the rearrangement of the remaining gives,
\begin{equation}
P=\frac{T}{2r_+}-\frac{1}{8\pi r_+^2}+\frac{Q^2}{8\pi a^2r_+^4} ~~,~~r_+=\left( \frac{3V}{4\pi a}\right)^{1/3}.
\label{eqstate}
\end{equation}
Therefore, equation (\ref{eqstate}) can be seen as $P=P(V,T)$, and hence the equation of state  in the extended space. The above form is called geometric equation of state.  Since, the variables are not having proper dimensions, we carry out adequate scaling as follows, 
\begin{equation}
\tilde{P}= \frac{\hbar c}{l_P^2}P~~,~~\tilde{T}=\frac{\hbar c}{k}T,
\end{equation} 
where $l_P$ is the Planck length. 
The functional dependence of the equation of state on $r_+$ tempts one to identify it with the van der Waals system at first sight. The comparison is complete by relating specific volume $v$ to the horizon radius $r_+$ as $v=2l_P^2r_+$. Finally, we arrive at the physical equation of state,

\begin{equation}
P=\frac{T}{v}-\frac{1}{2\pi v^2}+\frac{2Q^2}{\pi a^2v^4}.
\label{P monopole}
\end{equation}

The characteristic $P-v$ diagram (figure \ref{PVs}) is obtained from this equation (\ref{P monopole}) which has van der Waals like behaviour. The existence of three distinct regions with alternate negative, positive and negative slopes is also a beacon of critical behaviour. The negative slope regions correspond to a stable state of the system, whereas the positive regions are for unstable states since an increase in volume with pressure is physically meaningless. These unphysical regions can be handled via Maxwell construction where the oscillating part of the isotherm is replaced by a straight line. The Maxwell's equal-area law in extended phase space is,
\begin{equation}
\oint V dP =0.
\end{equation}

\begin{figure}[H]
    \subfigure[]
    {
        \includegraphics[width=0.5\textwidth]{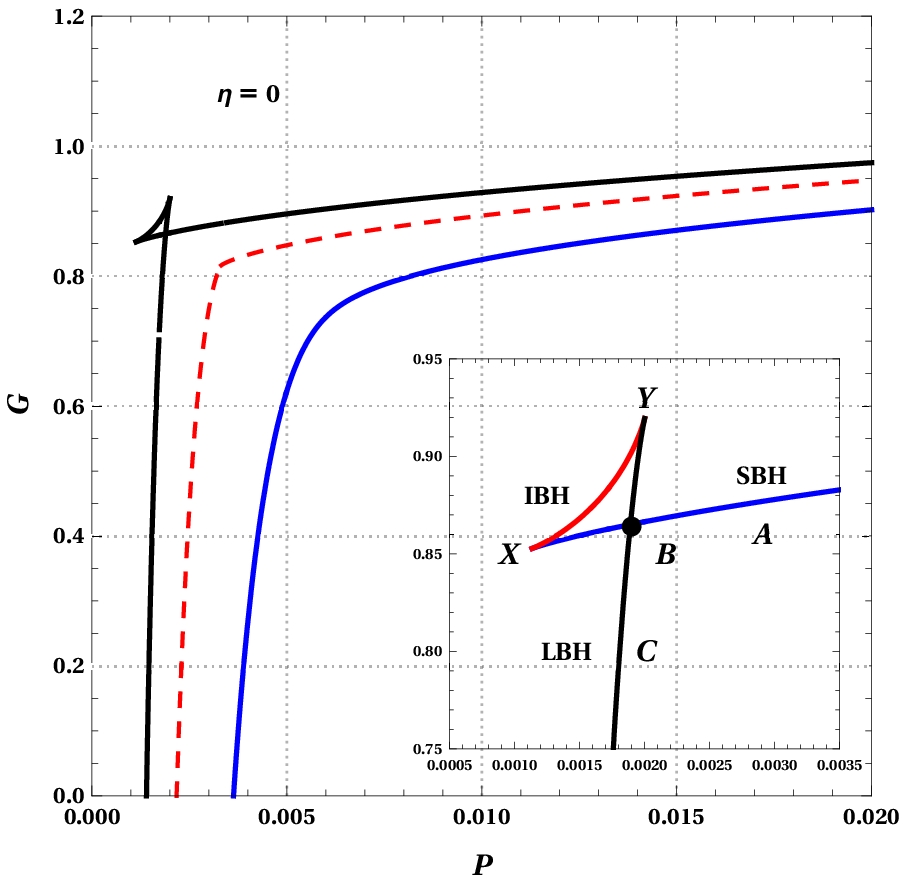}
        \label{GP1}
    }
    \subfigure[]
    {
        \includegraphics[width=0.5\textwidth]{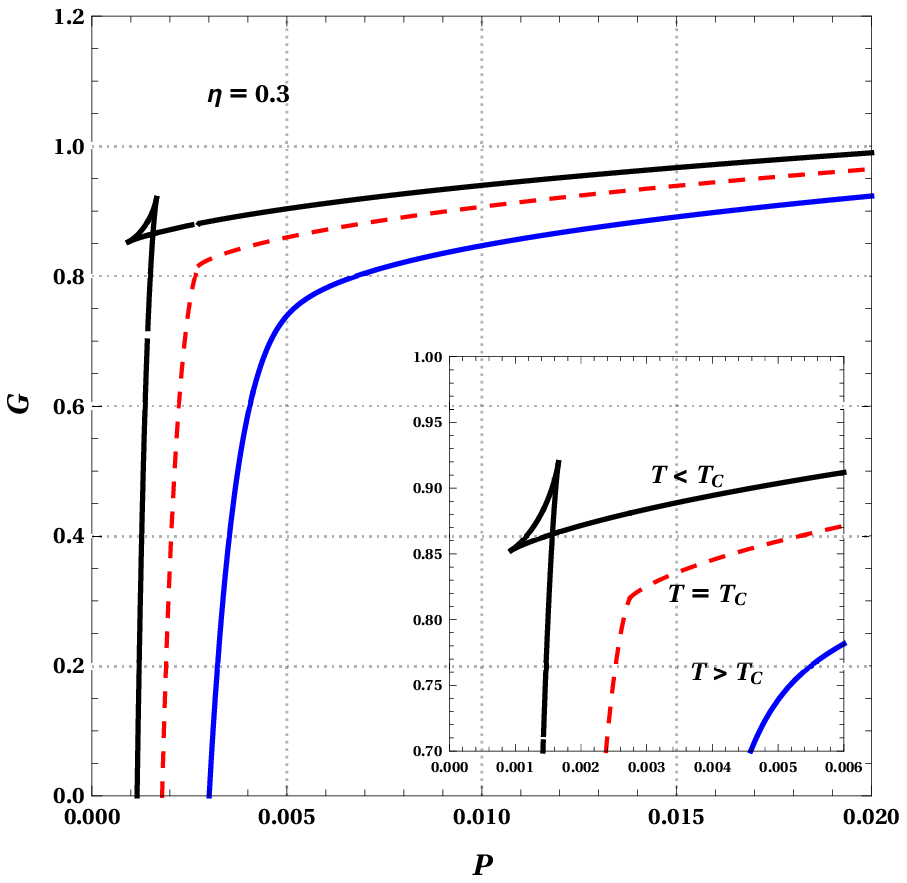}
        \label{GP2}
    }
        \subfigure[]
    {
        \includegraphics[width=0.5\textwidth]{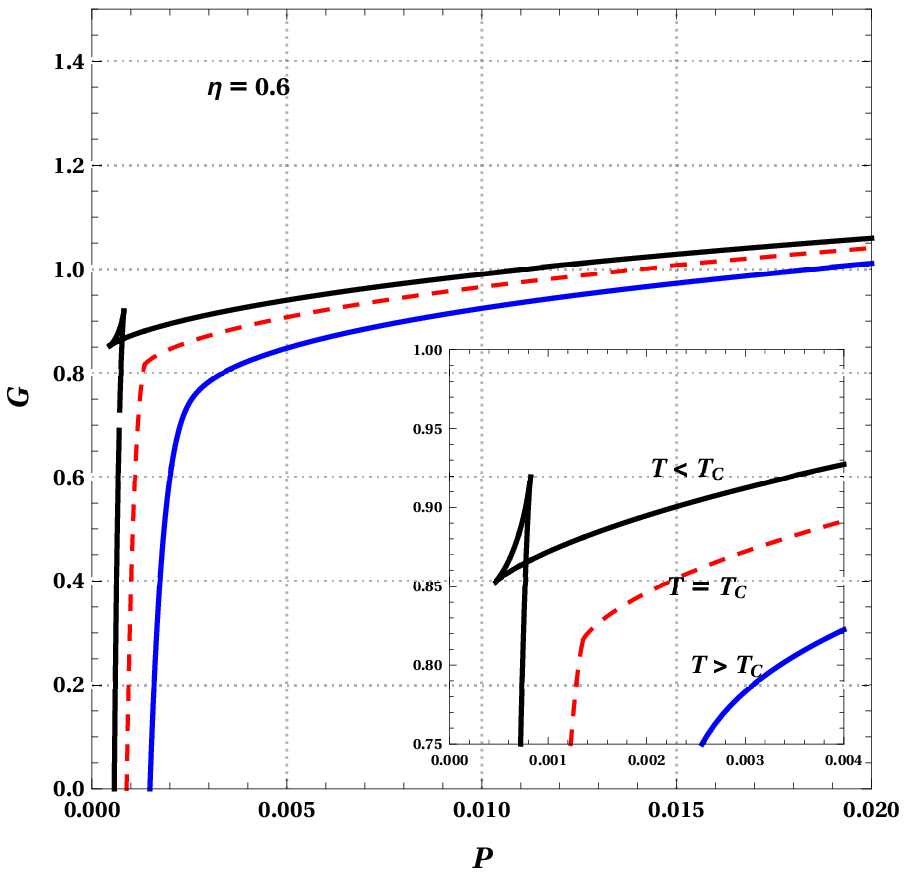}
        \label{GP3}
    }
    \subfigure[]
    {
        \includegraphics[width=0.5\textwidth]{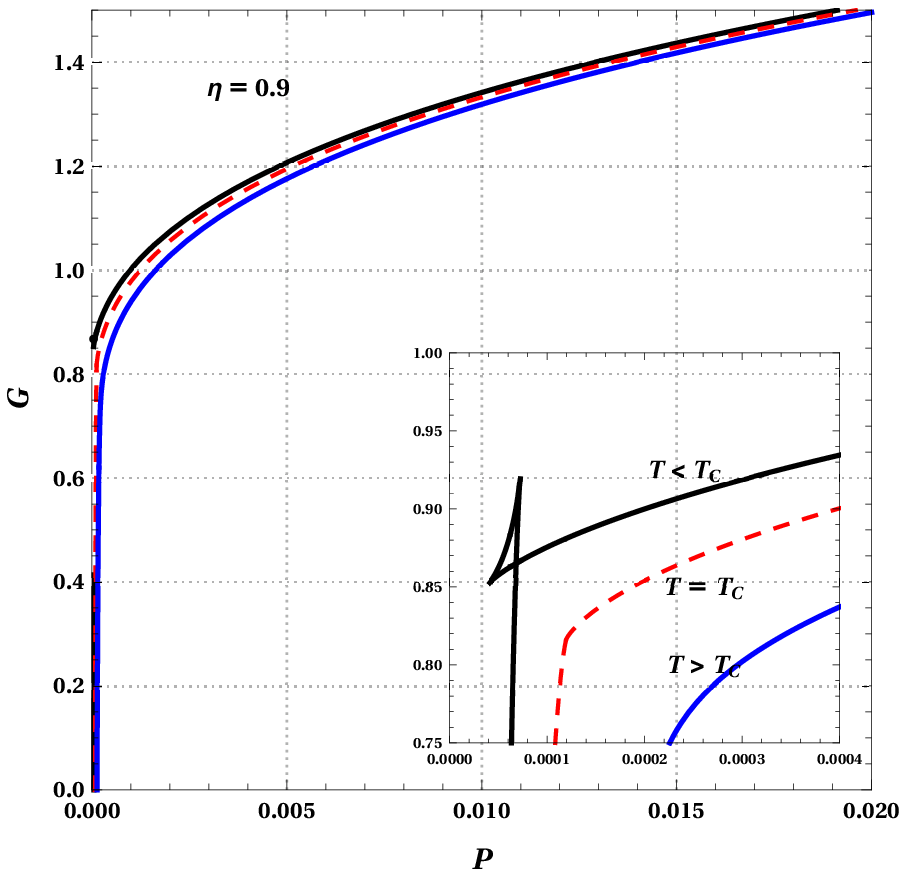}
        \label{GP4}
    }
\\
\caption{The swallow tail behaviour of Gibbs free energy and its variation with  $\eta$ in the extended phase space. The van der Waals like behaviour persists for all values of $\eta$. The diminishing behaviour is enlarged in inlets for close examination.   We have taken $Q=1$.}\label{GPplot1}
\end{figure}

It is well-established in the literature that, in a canonical ensemble where the charge is fixed, the asymptotically AdS black holes show a first-order phase transition analogous to van der Waals system terminating in a second-order critical point \citep{Chamblin1999}.  The effect of monopole parameter is seen in the series of plots (figure \ref{PVs}), where $\eta$ consistently suppresses the original behaviour of all the isotherms and bring them closer to the critical isotherm. Also, it appears as if $\eta$ removes the oscillating isotherms at the upper limit of its strength. This may be interpreted as the maximum value of $\eta$ destroys the van der Waals like nature of the charged black hole. But later we will see in Gibbs free energy plots that the inherent signature of criticality persists at least in the smaller form (figure \ref{GPplot1}) even at the maximum strength of $\eta$.

The critical parameters can be obtained by utilising the vanishing derivatives at the critical point, 
\begin{equation}
\left(\frac{\partial P}{\partial v}\right)_T=\left(\frac{\partial ^2 P}{\partial v^2}\right)_T=0.
\end{equation}
Which are
\begin{equation}
P_c=\frac{a^2}{96 \pi Q^2}~~,~~v_c=\frac{2\sqrt{6}Q}{a}~~,~~T_c=\frac{a}{3\sqrt{6}\pi Q}.
\label{extended ctitical}
\end{equation}
The presence of $\eta$ in these parameters once again validates our quest for its effect on thermodynamics of charged black holes. Compared to the RN-AdS black hole, critical quantities $P_c$ and $T_c$ decreases while $v_c$ increases with $\eta$ (since greater $\eta$ corresponds to smaller $a$).

As in classical thermodynamics, the critical behaviour of a system is more effectively represented by Gibbs free energy $G$. This is because of the thermodynamic potential $G$ measures global stability in an equilibrium process. In extended phase space, the total Euclidean action calculated for fixed $\Lambda$ is associated with Gibbs free energy \citep{Kubiznak2012}. One can obtain the Gibbs free energy by the Legendre transformation $G=M - T S$. In our case, it is calculated as follows,
\begin{equation}
G(P,T)= \frac{1}{4} a r_+ \left(1-\frac{8\pi Pr_+^2}{3}\right)+\frac{3 Q^2}{4 a r_+}.
\label{gibbs1}
\end{equation}

The behaviour of Gibbs free energy in terms of $P$ is illustrated in figure (\ref{GPplot1}). The $r_+$ in equation (\ref{gibbs1}) was replaced with $r_+(P,T)$ from equation of state (\ref{eqstate}).  For $T>T_c$, $G$ is single-valued and hence locally stable. It has a swallow tail nature below the critical temperature ($T<T_c$), which is a clear indication that there is a first-order phase transition in the system. This phase transition is between a small black hole (SBH) and a large black hole (LBH). The effect of $\eta$ in $G-P$  plots comes into play slowly when we increase its strength from zero to one. The swallow tail region (unstable states) persists for all values of $\eta$, but it's presence shrinks the tail to a smaller region. It appears as if the swallow tail disappears for larger monopole strength, but a more close observation falsifies this illusion (shown in the inset of figures).

The coexistence of large black hole and small black hole phases can be well depicted in a coexistence curve in $P-T$ plane. Along the coexistence curve, the black hole undergoes a first-order phase transition.  This can be achieved either from Maxwell's equal-area law or from Clausius-Clapeyron equation directly. Another elegant way of obtaining this curve is by exploiting the fact that the temperature and Gibbs free energy coincide for SBH (with radius $r=r_1$) and LBH (with radius $r=r_2$) along the coexistence curve \citep{Kubiznak2012, Mo:2016sel}.

\begin{figure}
\centering\includegraphics[width=0.45\textwidth]{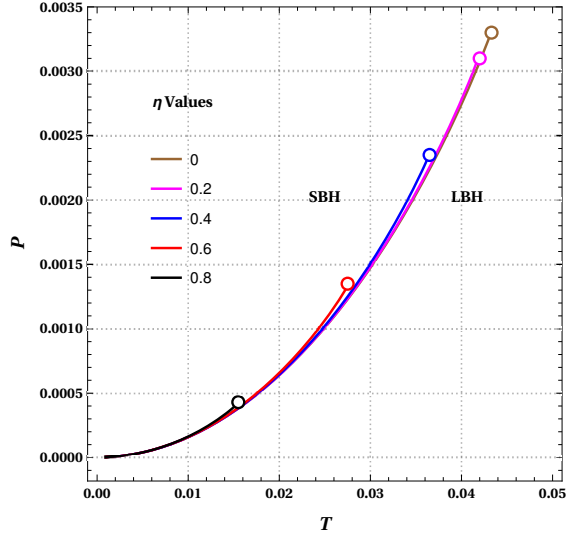}
\caption{Coexistent curve in extended phase space for different values of $\eta$. Coexistence line separates the LBH and SBH phases. The critical point where the first order transition terminates is marked as a circle.}\label{Coexteta_extended}
\end{figure}

We assert the following abbreviations for the simplification of calculation,
\begin{equation}
r_1+r_2=x, \quad r_1 r_2=y.
\end{equation}
The conditions which are mentioned earlier for coexistence curve lead us to the  set of three equations after some routine algebra. Equating Gibbs free energy on both sides,
\begin{equation}
\frac{1}{r_1}\left(9Q^2+3a^2r_1^2-8\pi a^2 Pr_1^4\right)=\frac{1}{r_2}\left(9Q^2+3a^2r_2^2-8\pi a^2 Pr_2^4\right)
\end{equation}
which reduces to 
\begin{equation}
3a^2y-8 \pi a^2Py(x^2-y)-9Q^2=0.
\label{G eqn} 
\end{equation}
Since the temperature on both sides are same, from equation of state we have,
\begin{equation}
T_0=\frac{1}{4\pi r_1}\left(1+\frac{3r_1^2}{l^2}-\frac{Q^2}{a^2r_1^2}\right)
\label{T01}
\end{equation}
\begin{equation}
T_0=\frac{1}{4\pi r_2}\left(1+\frac{3r_2^2}{l^2}-\frac{Q^2}{a^2r_2^2}\right).
\label{T02}
\end{equation}
Equating the  R.H.S of above two equations,
\begin{equation}
\frac{1}{r_1^3}\left(a^2r_1^2+8\pi P a^2r_1^4-Q^2\right)=\frac{1}{r_2^3}\left(a^2r_2^2+8\pi P a^2r_2^4-Q^2\right).
\end{equation}
which simplifies to
\begin{equation}
8\pi P a ^2y^3+Q^2(x^2-y)=a^2y^2.
\label{subtraction equation}
\end{equation}
Adding the equations (\ref{T01}) and (\ref{T02}),
\begin{eqnarray}
2T_0=\frac{1}{4\pi a^2}&\left[\frac{a^2r_1^2+8\pi Pa^2r_1^4-Q^2}{r_1^3} +\frac{a^2r_2^2+8\pi Pa^2r_2^4-Q^2}{r_2^3}\right], 
\end{eqnarray}
and simplifying,
\begin{equation}
8\pi T_0 a^2 y^3=a^2y^2x+8\pi Pa^2y^3x-Q^2x(x^2-3y).
\label{addition equation}
\end{equation}
The equations (\ref{G eqn}), (\ref{subtraction equation}) and (\ref{addition equation}) are solved for $P-T$ plane and the result is displayed in figure (\ref{Coexteta_extended}). The curve is quite similar to the van der Waals system.

In the coexistence curve, an increase in $\eta$ reduces the region for the coexistence. This is another proof for the aforementioned argument that the monopole term is a hindrance for the critical behaviour. This argument is based on two defining features of the coexistence curve. Firstly, crossing the curve in any way, stands for a first-order phase transition. Secondly, the termination of the curve is at a second-order transition point. Lowering of termination point is related to the fact that $P_c$ and $T_c$ reduces with increasing $\eta$. Smaller region of coexistence implies a smaller range of pressure and temperature, which gives phase transition.

\subsection*{Critical Exponents}
Here we compute the critical exponents $\alpha , \beta , \gamma, \delta$ for the black hole with monopole term in extended phase space. These universal exponents describe the behaviour of response functions near the critical point. The exponent $\alpha$ is related to the specific heat, $\beta$ characterizes the order parameter, $\gamma$ characterises the isothermal compressibility and $\delta$ is the measure of flatness of the critical isotherm. First we investigate the behaviour of specific heat $C_V$, which can be obtained from the free energy,
\begin{equation}
F=G-PV=\frac{1}{2}\left(ar_+-2\pi T a r_+^2+\frac{Q^2}{ar_+}\right).
\end{equation}
From this entropy can be calculated as,
\begin{equation}
S(T,V)=-\left( \frac{\partial F}{\partial T}\right) _V=\pi a r_+^2
\label{just above}
\end{equation}
followed by the inference that $C_V=0$, since there is no dependence on temperature $T$ in equation (\ref{just above}). Since  $C_V \propto |t|^\alpha $ the exponent $\alpha =0$.

The law of corresponding states is obtained by writing the equation of state in terms of the reduced thermodynamic variables,
\begin{equation}
p=\frac{P}{P_C}, \quad \nu=\frac{v}{v_C}, \quad \tau =\frac{T}{T_C}
\end{equation}
the relevant among them are also written in a way how much they differ from critical points as $\nu=1+\omega$ and  $\tau=1+t$. Using these and the expressions for critical values (equation \ref{extended ctitical}), the equation of state (\ref{P monopole}) reduces to 
\begin{equation}
p=\frac{8}{3}\frac{\tau}{\nu}-\frac{2}{\nu^2}+\frac{1}{3\nu^4}.
\label{reducedeqn}
\end{equation}
This equation (\ref{reducedeqn}) is not altered by the presence of global monopole. Therefore, the remaining calculations on critical exponents are identical to charged AdS black hole \citep{Kubiznak2012}. Expanding the above (equation \ref{reducedeqn})  around the critical point we get
\begin{equation}
p=1+\frac{8}{3}t-\frac{8}{9}t\omega-\frac{4}{81}\omega ^3+O(t\omega ^2, \omega ^4).
\label{law of corresponding}
\end{equation}
Differentiating this with respect to $\omega$ and using Maxwell's equal area law we obtain,
\begin{equation}
p=1+\frac{8}{3}t-\frac{8}{9}t\omega _l-\frac{4}{81}\omega _l ^3=1+\frac{8}{3}t-\frac{8}{9}t\omega _s-\frac{4}{81}\omega _s ^3
\end{equation}
and
\begin{equation}
0=\int _{\omega _l} ^{\omega _s}\omega (6t+\omega ^2)d\omega .
\end{equation}
The above two equations have unique solution $\omega _s=-\omega _l=3\sqrt{-2t}$.
Now we can calculate,
\begin{equation}
\tilde{\eta} = V_C(\omega _l-\omega _s)=2V_C\omega _l=6V_C \sqrt{-2t}.
\end{equation}
Since $\tilde{ \eta } \propto |t|^\beta $ we have $\beta =1/2$.
Differentiating equation (\ref{law of corresponding}) with respect to $V$ and inverting,
\begin{equation}
\left . \frac{\partial V}{\partial T}\right|_T\propto -\frac{9}{8}\frac{V_C}{T_C}\frac{1}{t}.
\end{equation}
And hence 
\begin{equation}
\kappa _T=-\frac{1}{V}\frac{\partial V}{\partial T}\propto \frac{1}{t} 
\end{equation}
From $\kappa _T \propto |t|^{-\gamma}$ we get $\gamma =1$.
The remaining critical exponent $\delta$ is obtained by setting $t=0$ in equation (\ref{law of corresponding}), which is the shape of the critical isotherm,
\begin{equation}
p-1=-\frac{4}{81}\omega ^3
\end{equation}
Since $p-1 \propto |\omega |^\delta$, $\delta=3.$
All the critical exponents are unaffected by the presence of $\eta$ and exactly matches with that of van der Waals system as in the case of RN-AdS black hole. The critical exponents must satisfy the universal scaling laws,
\begin{equation}
\alpha + 2\beta +\gamma=2 \quad , \quad \gamma =\beta (\delta -1).
\end{equation}
These are satisfied in our case. This means that the SBH-LBH phase transition is analogous to van der Waals liquid-gas system and belongs to the same universality class.  

\section{Thermodynamics in Alternate Phase Space}
\label{Phase alternate}
\begin{figure}[H]
    \subfigure[]
    {
        \includegraphics[width=0.5\textwidth]{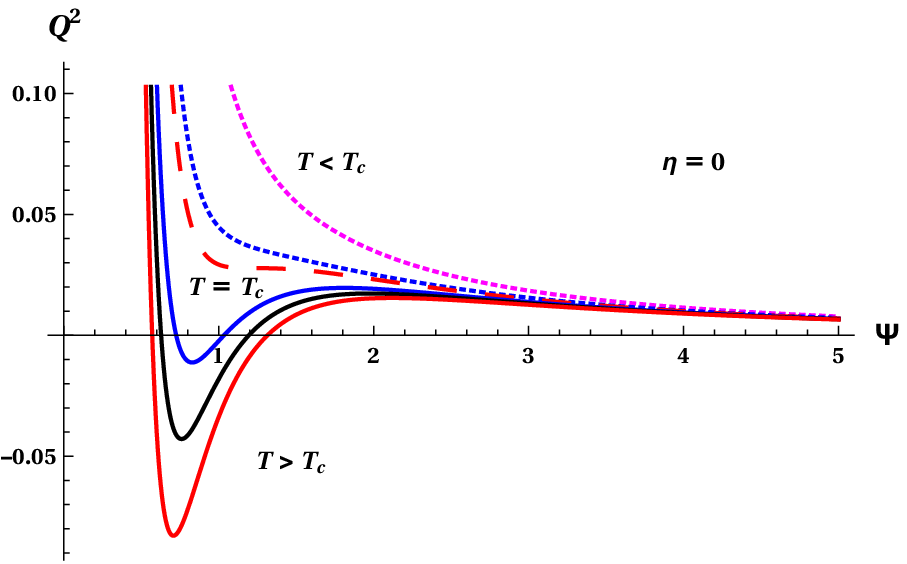}
        \label{Q1}
    }
    \subfigure[]
    {
        \includegraphics[width=0.5\textwidth]{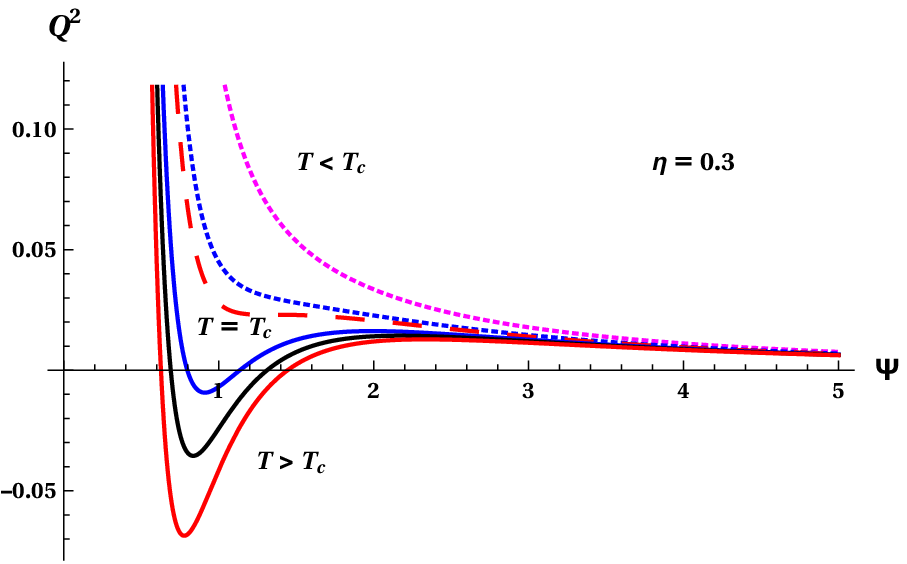}
        \label{Q2}
    }
        \subfigure[]
    {
        \includegraphics[width=0.5\textwidth]{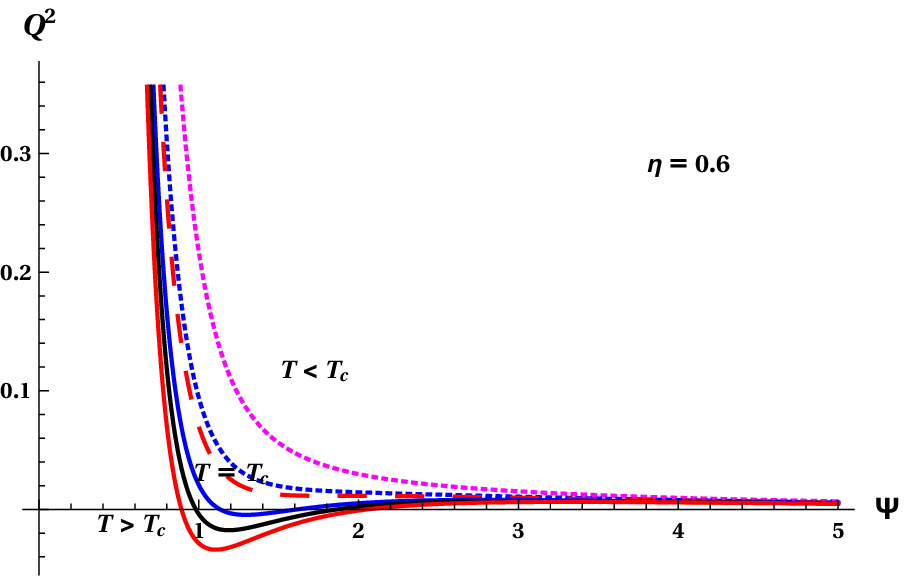}
        \label{Q3}
    }
    \subfigure[]
    {
        \includegraphics[width=0.5\textwidth]{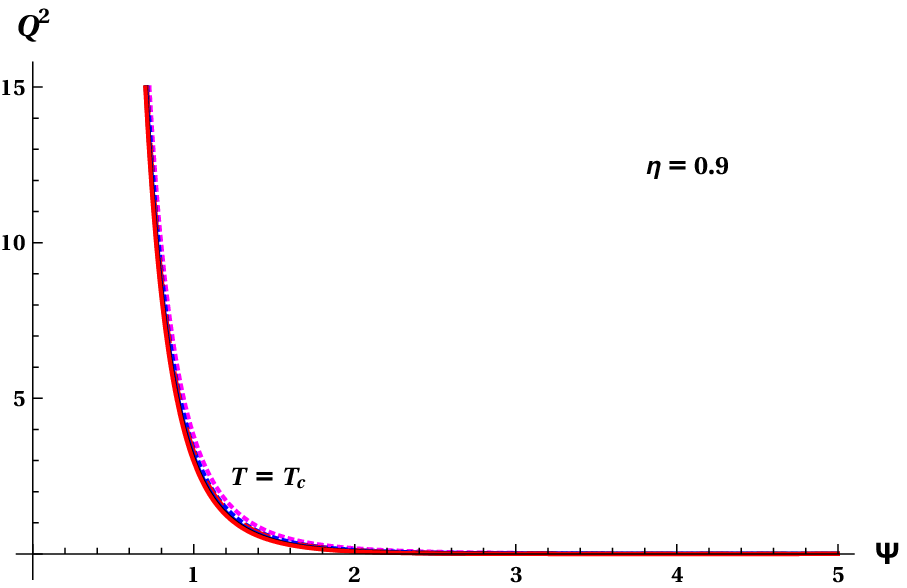}
        \label{Q4}
    }
\\
\caption{The $Q^2-\Psi$ isotherms for charged AdS black hole with a global monopole. The plots are shown for different values of $\eta$ for the case of $l=1$. The oscillatory behaviour is displayed for temperatures above $T_C$. The figures illustrates the effect of $\eta$, where increasing its strength reduces the oscillatory behaviour. }\label{Q2psiplot}
\end{figure}

In this section, we study the thermodynamics of the black hole in an alternate approach. Here we identify the square of the charge $Q^2$ with fluid pressure and keep the cosmological constant fixed. The mathematically independent conjugate variable is chosen to be inverse of specific volume, $\Psi =1/v$.   We write the equation of state in terms of $Q^2$ followed by the study of the critical behaviour of isotherms in $Q^2-\Psi$ plane. We also investigate the effect of global monopole in the phase transition in this alternate approach for different values of $\eta$.
First law takes a modified form in alternative space as follows,
\begin{equation}
dM=TdS+\Psi dQ^2+VdP
\label{smarr}
\end{equation}
with the corresponding Smarr formula,
\begin{equation}
M=2(TS+\Psi Q^2-VP).
\end{equation}
From equations (\ref{eq9}) and (\ref{smarr}),
\begin{equation}
\Psi=\frac{\partial M}{\partial Q^2}=\frac{1}{2ar_+}.
\label{Psi}
\end{equation}
We begin by writing the equation of state as $Q^2(\Psi, T)$. So, we rearrange the expression for Hawking temperature (equation \ref{eqT}) as, 
\begin{equation}
Q^2=a^2\left[ r_+^2+\frac{3r_+^4}{l^2}-4\pi r_+^3T\right].
\label{eos alternate}
\end{equation}
Inserting equation (\ref{Psi}) into this we obtain the equation of state,
\begin{equation}
Q^2=\frac{1}{4} \left[\frac{1}{ \psi ^2}+\frac{3}{4 a^2 l^2 \psi ^4}-\frac{2\pi  T}{ a \psi ^3}\right].
\label{eos alt}
\end{equation}
From the equation of state one can obtain $Q^2-\Psi$ isotherms which are illustrated in figure(\ref{Q2psiplot}). These isotherms are analogous to van der Waals isotherms showing a first order phase transition between a SBH and LBH phases. Contrast to $P-v$ isotherms, here we have oscillating isotherms for $T>T_C$. In $Q^2-\Psi$ isotherms the intermediate region has positive slope $(\partial Q^2/\partial \Psi >0)$ which corresponds to the unstable states and the other two are negative slope regions representing stable states. The physically irrelevant parts of the graph, i.e, the positive slope region and negative $Q^2$ region are eliminated by Maxwell construction. In the alternate phase space Maxwell's equal area law reads as 
\begin{equation}
\oint \Psi  d Q^2 =0.
\end{equation}
It is clear from the figure (\ref{Q2psiplot}) that $\eta$ changes the quantitative behaviour of the isotherms. As $\eta$ approaches to unity all isotherms tends towards the critical isotherm. This is similar to the result we obtained in extended phase space for $P-v$ isotherms.

In the critical isotherm there is an inflection point at which we can define the critical point,
\begin{equation}
\left. \frac{\partial Q^2}{\partial \Psi }\right| _{T_C}=0, \quad \left. \frac{\partial ^2 Q^2}{\partial \Psi ^2 }\right| _{T_C}=0.
\end{equation}
This gives the following critical parameters,
\begin{equation}
T_C=\frac{1}{\pi l} \sqrt{\frac{2}{3}}, \quad Q_C^2=\frac{l^2}{36} a^2, \quad \Psi _C=\frac{1}{a} \sqrt{\frac{3}{2l^2}}
\end{equation}
Unlike in the case of extended phase space, the critical temperature is independent of $a$. The product of critical parameters,
\begin{equation}
\rho _C=Q_C^2 T_C \Psi _C=\frac{a}{36\pi}
\end{equation}
In the limiting case $\eta =0 \Rightarrow a=1$ this reduces to the universal constant $\rho _C=1/36\pi$ which is consistent with the result obtained in RN-AdS black hole \citep{Dehyadegari2017}.

\begin{figure}[H]
    \subfigure[]
    {
        \includegraphics[width=0.5\textwidth]{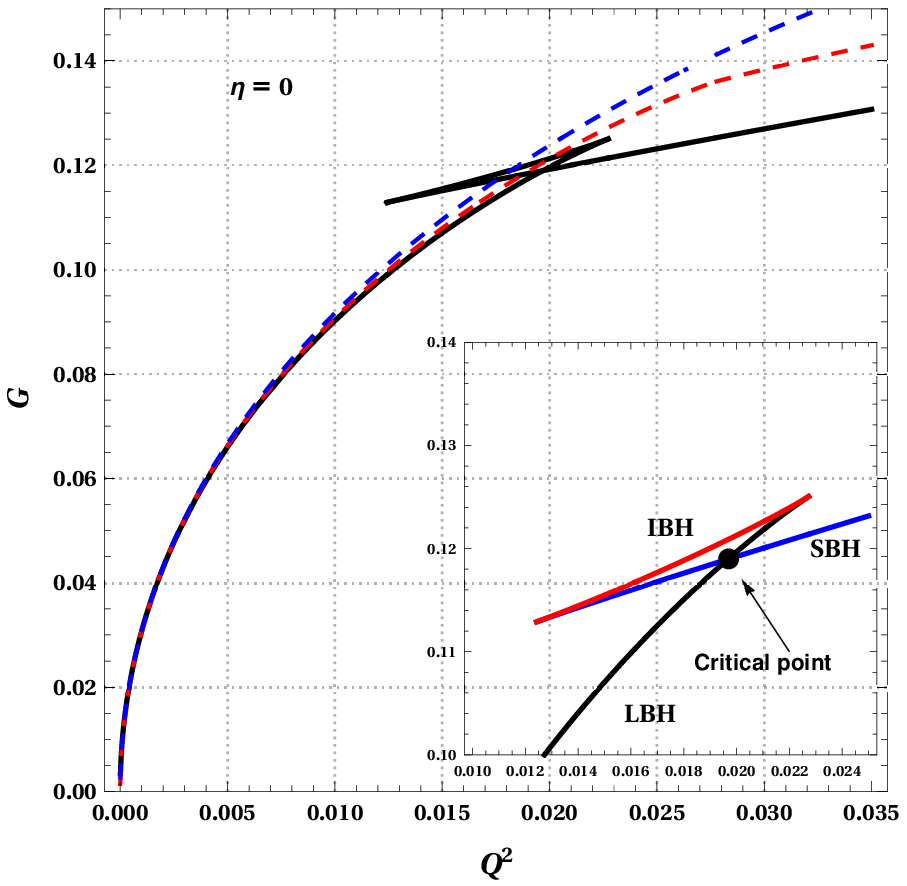}
        \label{G1}
    }
    \subfigure[]
    {
        \includegraphics[width=0.5\textwidth]{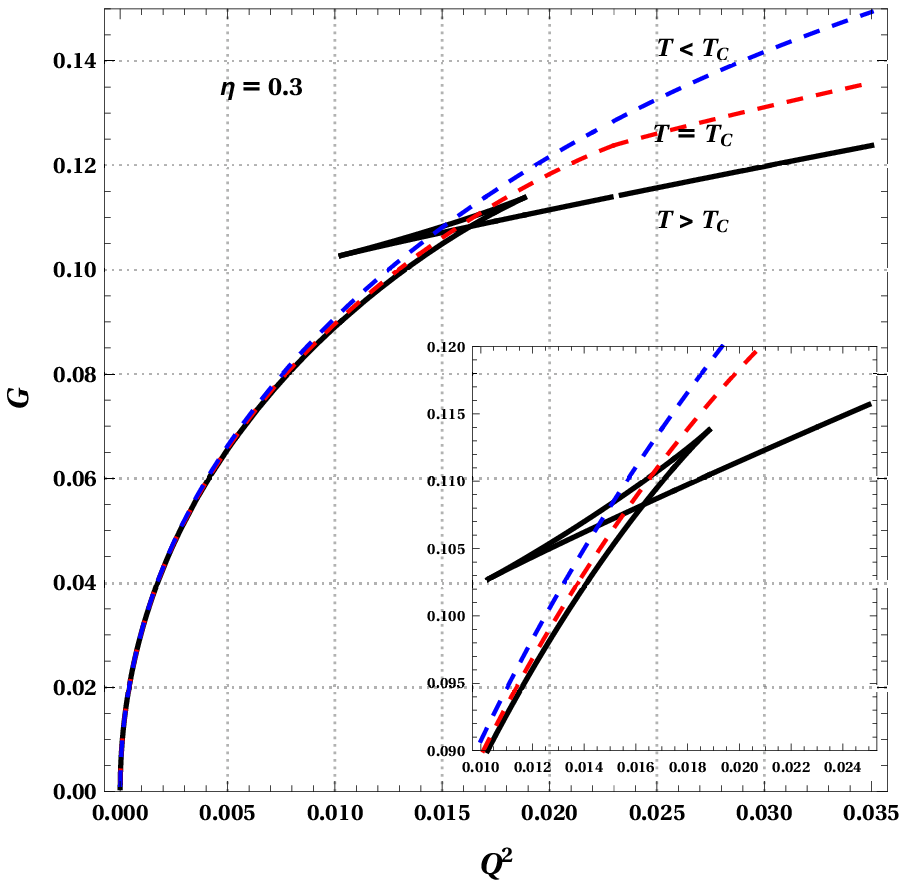}
        \label{G2}
    }
        \subfigure[]
    {
        \includegraphics[width=0.5\textwidth]{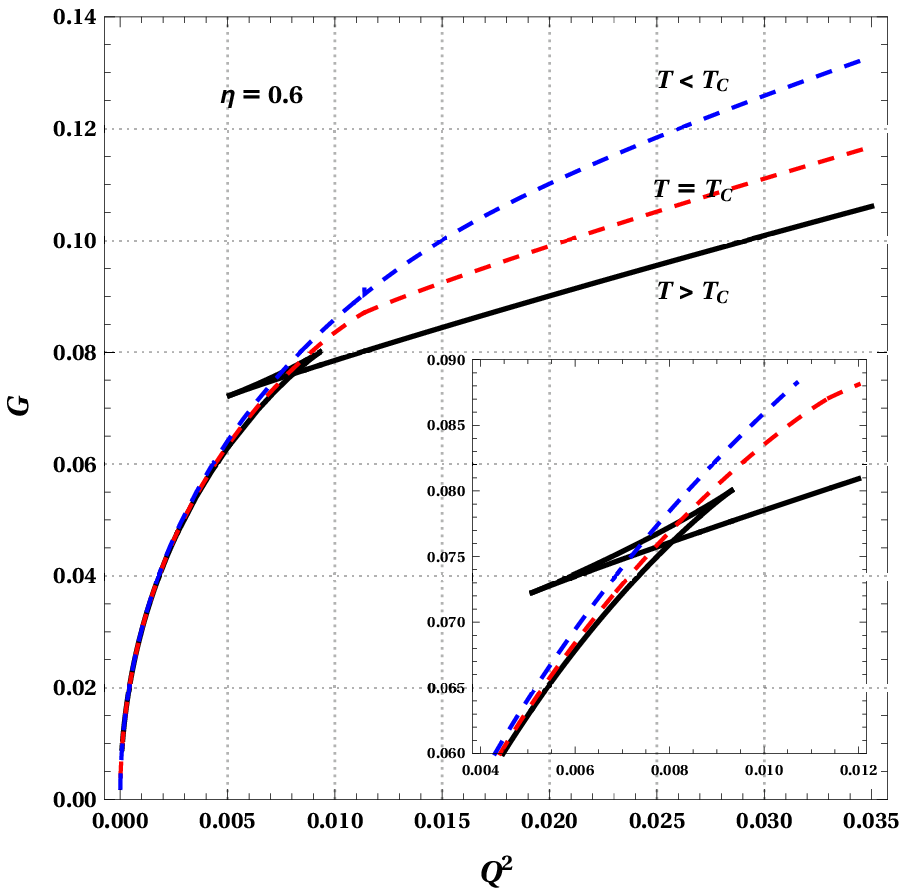}
        \label{G3}
    }
    \subfigure[]
    {
        \includegraphics[width=0.5\textwidth]{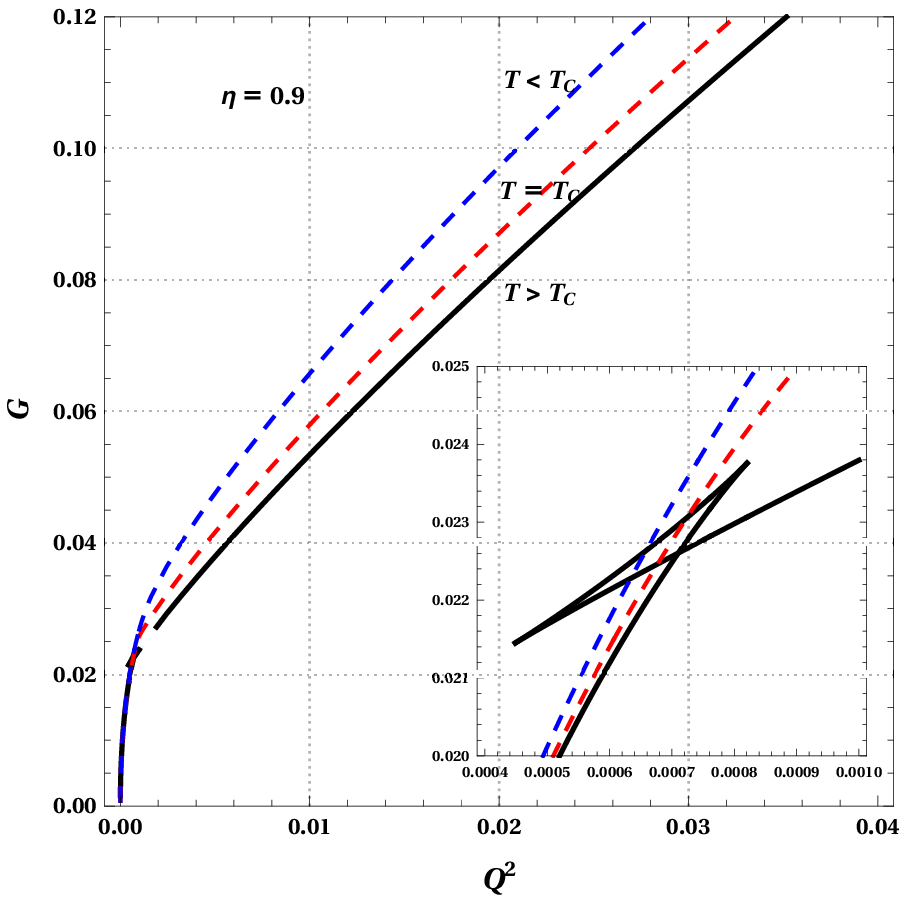}
        \label{G4}
    }
\\
\caption{Gibbs free energy plots for different values of $\eta$. In the inlet of figure (\ref{G1}) the SBH, LBH and IBH branches are shown. The critical point where the first order transition takes place also marked. In all these figures for $T>T_C$, there is a swallow tail behaviour. When $\eta$ increases this swallow tail diminishes and it is enlarged in inlets.}\label{GQ2plot1}
\end{figure}

To study more about phase transition we need Gibbs free energy, which is obtained from the Legendre transformation $G=M-TS$,
\begin{equation}
G(Q^2,T)= \frac{1}{4} a r_+ \left(1-\frac{r_+^2}{l^2}\right)+\frac{3 Q^2}{4 a r_+}.
\end{equation}
In this expression $r_+$ is understood as $r_+(Q^2,T)$ from equation (\ref{eos alternate}). The behaviour of Gibbs free energy in terms of $Q^2$ for different values of $\eta$ are shown in figure (\ref{GQ2plot1}). For $T>T_C$ there exist three branches with discontinuities. The branches $A-B$, $B-C$ and $X-Y$ correspond to a small black hole (SBH), large black hole (LBH) and intermediate black hole (IBH) phases respectively. For higher $Q^2$ values an SBH phase is preferred because the free energy for SBH is larger compared to LBH. In the smaller $Q^2$ region LBH phase is favoured. In between LBH and SBH phases, there is a coexistence point where two phases co-exist. While crossing this critical point, a first-order phase transition occurs with the release of finite latent heat. However, these SBH and LBH branches do not terminate at the coexistence point but continue a little further, which represents metastable states. These metastable states stay for a small interval of time. In classical thermodynamics, these states are called superheating and supercooling phases. LBH and SBH branches are joined together via an unstable phase IBH. This IBH phase is separated by a finite jump from the other two. At $T=T_C$, $G$ is single-valued, continuous but non-analytic, corresponding to a second-order phase transition between SBH and LBH. For $T<T_C$ the triangle $BXY$ disappears, i.e there is no sharp distinction between SBH and LBH below the critical temperature. Below that temperature one cannot say where the system starts being LBH or stops being SBH. As $\eta$ increases from zero to unity, the swallow tail approaches the origin, but the multivaluedness of $G$ for $T>T_C$ remains. The IBH phase is reduced to a smaller region with a larger $\eta$ strength.

After studying the behaviour of Gibbs free energy it is necessary to explore the coexistence curve for SBH and LBH. Along the coexistence curve SBH and LBH phases coexist with the same temperature and Gibbs free energy. As in extended space calculations we take following abbreviations,
\begin{equation}
r_1+r_2=x, \quad r_1 r_2=y.
\end{equation}

\begin{figure}
\centering\includegraphics[width=0.45\textwidth]{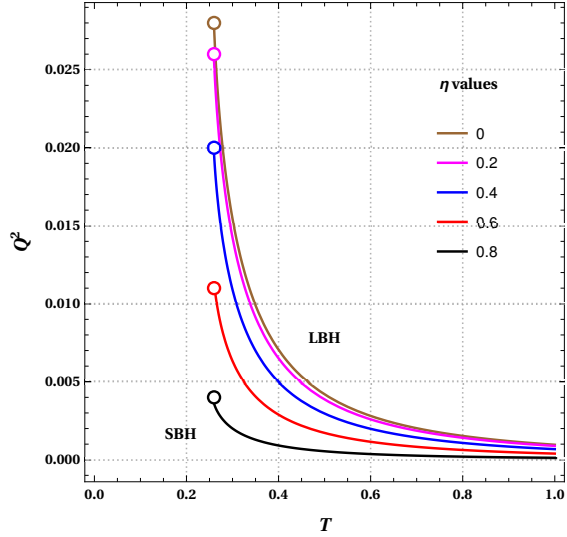}
\caption{Coexistent curve in alternate phase space for different values of $\eta$. The coexistence phase exists only for $T>T_C$. Critical points are labeled with a circle.}\label{Coexteta}
\end{figure}

Equating the Gibbs free energy on both sides (we set $l=1$),
\begin{equation}
\frac{1}{r_1}\left[3Q^2+a^2(r_1^2-r_1^4)\right]=\frac{1}{r_2}\left[3Q^2+a^2(r_2^2-r_2^4)\right].
\end{equation}
In terms of $x$ and $y$ this reduces to,
\begin{equation}
3Q^2+a^2 y (x^2-y)=a^2 y.
\label{Equate G}
\end{equation}
The temperature on both sides are given by equations (\ref{T01}) and (\ref{T02}). Equating the R.H.S of those two equations we have,
\begin{equation}
\frac{1}{r_1^3}\left(a^2r_1^2+3a^2r_1^4 -Q^2\right)=\frac{1}{r_2^3}\left(a^2r_2^2+3a^2r_2^4 -Q^2\right).
\end{equation}
Which reduces to 
\begin{equation}
3a^2y^3+Q^2(x^2-y)=a^2y^2.
\label{Equate T}
\end{equation}
Adding the  equations (\ref{T01}) and (\ref{T02}) we get,
\begin{equation}
2T_0=\frac{1}{4\pi a^2}\left[ \frac{a^2r_1^2+3a^2r_1^4-Q^2}{r_1^3}+\frac{a^2r_2^2+3a^2r_2^4-Q^2}{r_2^3}\right].
\end{equation}
Changing the variables to $x$ and $y$,
\begin{equation}
8 \pi  T_0a^2 y^3=a^2y^2 x +3a^2y^3x -Q^2x(x^2-3 y).
\label{Add T}
\end{equation}
The coexistence line is obtained by solving the equations (\ref{Equate G}), (\ref{Equate T}) and (\ref{Add T}) for $Q^2-T$ plane. The coexistence curve in alternate phase space is shown in figure (\ref{Coexteta}). Since $T_C$ is independent of $\eta$, all curves begins from the same point for all values of $\eta$. Yet, $\eta$ contributes to the coexistence curve, which is evident from the figure (\ref{Coexteta}), with reduction of coexistence region on increasing $\eta$. (Note that $Q^2_C$ depends on $\eta$).  

\subsection*{Critical exponents}
As we mentioned earlier, the critical exponents characterize the behaviour of thermodynamic functions near the critical point. The critical exponents in alternate phase space are calculated as follows. We start by defining the reduced thermodynamic variables
\begin{equation}
\Psi _r\equiv\frac{\Psi }{\Psi _C}, \quad Q _r^2\equiv\frac{Q^2}{Q^2 _C},  \quad T _r\equiv\frac{T}{T _C}. 
\label{reduced}
\end{equation}

The reduced variables are then written as $T_r=1+t$, $\Psi _r=1+\psi $ and $Q_r^2=1+\mathcal{Q}$, where $t, \psi $ and $\mathcal{Q}$ are the deviation from the critical point. To find the exponent $\alpha$ we consider the entropy as a function of temperature $T$ and $\Psi =1/(2ar_+)$,
\begin{equation}
S=S(T,\Psi)=\frac{\pi}{4a\Psi ^2}
\end{equation}
which is independent of temperature $T$. Therefore the specific heat at fixed $\Psi$ vanishes,
\begin{equation}
C_\Psi =\left. T\frac{\partial S}{\partial T}\right|_\Psi =0.
\label{alpha}
\end{equation}
The vanishing $C_\Psi$ implies that $\alpha =0$, since  $C_\Psi \propto |t|^\alpha $.
The equation of state (\ref{eos alternate}) can be written in terms of reduced thermodynamic variables (equation \ref{reduced}) as follows, 
\begin{equation}
Q_r^2=\frac{6}{\Psi _r^2}+\frac{3}{\Psi _r^4}-\frac{8T_r}{\Psi _r^3}.
\label{reduced eos}
\end{equation}
This law of corresponding states is independent of $a$ and $l$. Therefore, the results using this expression will be identical to that of RN-AdS black hole \citep{Dehyadegari2017}. Ignoring the higher order terms near the critical point, equation (\ref{reduced eos}) takes the following form
\begin{equation}
\mathcal{Q} =-8t+24 t \psi - 4\psi ^3+o\left( t\psi ^2,\psi ^4\right).
\label{rho}
\end{equation}
The Maxwell's equal area law and the derivative of equation (\ref{rho}) with respect to $\psi$ at a fixed $t>0$ yields,
\begin{equation}
\mathcal{Q}=-8t+24 t \psi _l - 4\psi _l^3=-8t+24 t \psi _s- 4\psi _s ^3
\end{equation}
\begin{equation}
0=\Psi _C\int _{\psi _l} ^{\psi _s}\psi (24t-12\psi ^2)d\psi .
\end{equation}
Keeping the analogy between the volumes of liquid and gas system, the inverse of specific volumes of LBH and SBH are identified with $\psi _l$ and $\psi _s$. The above equations have the nontrivial solution
\begin{equation}
\psi _s=-\psi _l=\sqrt{6t}.
\end{equation}
This gives the order parameter near the critical point,
\begin{equation}
|\psi _s-\psi _l|=2\psi _s=2\sqrt{2}t^{1/2}\Rightarrow \beta =1/2.
\end{equation}
Now, let's consider the critical exponent $\gamma$, which determines the functional behaviour of $\chi _T$ near the critical point,
\begin{equation}
\chi _T=\left. \frac{\partial \Psi}{\partial Q^2}\right|_T \propto |t|^{-\gamma}. 
\end{equation}
From equation (\ref{rho}) we get,
\begin{equation}
\chi _T=\frac{\Psi _C}{24Q_C^2t}\Rightarrow\gamma =1.
\end{equation}
And finally considering $\mathcal{Q}$ at $t=0$ we have $\mathcal{Q}=-4\psi ^3\Rightarrow\delta =3 $ which depicts the shape of the critical isotherm as in extended phase space. All the critical exponents obtained are same as in extended phase space, i.e., it matches with van der Waals system. 

Now that the black hole shows phase transitions, one may be interested in the microscopic structure of that black hole. Before investigating it, we make some remarks about extended and alternate phase spaces. While observing the thermodynamics of black hole in extended phase space we could make the correspondence to van der Waals system. The same inference in alternate space affirms that this new approach is as good as the old one. However, we note that taking $Q^2$ as a thermodynamic variable than cosmological constant makes more sense physically for a charged black hole. In this way, we can attribute the critical behaviour to the charge of the black hole. The additional parameter $\eta$ do not change this notion.

\newpage
\section{Thermodynamic Geometry and Microscopic Structure}
\label{TD Geometry}
Finally, we investigate the microscopic structure of the black hole using Ruppeiner geometry. This is the explicit form of the inherent geometric structure of thermodynamics. In thermodynamic fluctuation theory probability of any fluctuation is proportional to the exponential of the invariant line element defined in the thermodynamic information geometry. The metric corresponding to this invariant Ruppeiner line element is defined as
\begin{equation}
g_{ij}=-\frac{\partial ^2S}{\partial X^i \partial X^j},
\end{equation}
where $X^i$ is any other thermodynamic extensive variable. Usually, for calculation purpose, the above  metric is written in Weinhold form,
\begin{equation}
g_{ij}=\frac{1}{T}\frac{\partial ^2M}{\partial Y^i \partial Y^j}.
\end{equation}
The scalar curvature obtained from this metric, known as Ruppeiner invariant $R$, contains the information about the first-order phase transition. This is shown in $R-T$ plane, where the multivaluedness of $R$ displays the coexistence of LBH and SBH, which in turn corresponds to the multivaluedness of thermodynamic potentials. The curvature scalar $R$ gives insight into the microscopic structure and the Black hole molecular interaction at the horizon. More precisely, the magnitude of $R$ is the measure of the strength of the interaction and its sign is the indicator of the nature of microscopic interaction. In the context of quantum interactions, it is shown that negative $R$ indicates an attractive Bosonic interaction and positive $R$ indicates the repulsive Fermi interactions \citep{Mrugala}. Whereas, for anyon gas, both attractive and repulsive statistical interactions are possible. In this case, for \emph{Bose-like attractive} interactions, $R$ is positive and for \emph{Fermi-like repulsive} interactions, $R$ is negative \citep{Mirza2008}.  If  $R=0$ then there is no interaction like ideal gas system. However, in the framework of black hole thermodynamics, a proper physical and analytical interpretation is not yet well established. This ambiguous interpretation persists because of the sign of $R$, which changes for different ensembles \citep{Sahay:2010wi}. The difficulty in addressing this question stems from the lack of knowledge about the constituents of the black hole. Nevertheless, it is accepted as a convention that $|R|$ is the measure of the stability of the black hole.

\subsection{Extended Phase Space Geometry}

The curvature scalar in extended phase space is calculated by taking the coordinates $Y^i=(S, P)$ and performing usual Riemannian geometry calculation via the metric components,
\begin{equation}
R=\frac{2 \pi  Q^2-a S}{8 P S^3 +a S^2-\pi  Q^2 S}.
\end{equation}
As we discussed earlier, the curvature scalar $R$ is having branches in the vicinity of the first-order phase transition. Along the transition line, the branches of $R$ corresponds to LBH and SBH, and they join at the critical point. We obtain analytical expressions for those two branches by adopting the following method. Similar analysis of $R-\chi$ plots for different spacetimes in various scenarios are discussed in the literature \citep{Guo2019, Miao2017, Zangeneh2017, Miao2019a, Chen2019, Du2019}.

Under Maxwell construction in $P-V$ plane, two points, $(P_0,V_1)$ and $(P_0,V_2)$ in the isotherm $T=T_0$ are joined by a line. These two points corresponds to SBH and LBH with event horizon $r_1$ and $r_2$ at first order transition point. From Maxwell's equal area law in extended phase space,
\begin{equation}
P_0(V_2-V_1)=\int _{r_1}^{r_2} PdV.
\end{equation}
With $r_1=xr_2$ this reduces to 
\begin{equation}
2P_0=\frac{1}{(1+x+x^2)}\left[ \frac{3}{2}\frac{T_0(1+x)}{r_2}-\frac{3}{4\pi r_2^2}+\frac{3Q^2}{4\pi a^2 r_2^4 x}\right].
\label{Maxarea}
\end{equation}
Pressure at those two points can be written as
\begin{equation}
P_0=\frac{T_0}{2r_1}-\frac{1}{8\pi r_1^2}+\frac{Q^2}{8\pi a^2 r_1^4}
\label{P01ext}
\end{equation}
and 
\begin{equation}
P_0=\frac{T_0}{2r_2}-\frac{1}{8\pi r_2^2}+\frac{Q^2}{8\pi a^2 r_2^4}.
\label{P02ext}
\end{equation}
Adding (\ref{P01ext}) and (\ref{P02ext}) we get 
\begin{equation}
2P_0=\frac{T_0(1+x)}{2r_2x}-\frac{(1+x^2)}{8\pi r_2^2x^2}+\frac{Q^2(1+x^4)}{8\pi a^2 r_2^4x^4},
\label{addExt}
\end{equation}
and equating R.H.S. of the same set of equations,
\begin{equation}
0=T_0-\frac{1}{4\pi}\frac{(1+x)}{r_2x}+\frac{Q^2(1+x^2)(1+x)}{4\pi a^2r_2^3x^3}.
\label{subExt}
\end{equation}

\begin{figure}[H]
    \subfigure[]
    {
        \includegraphics[width=0.5\textwidth]{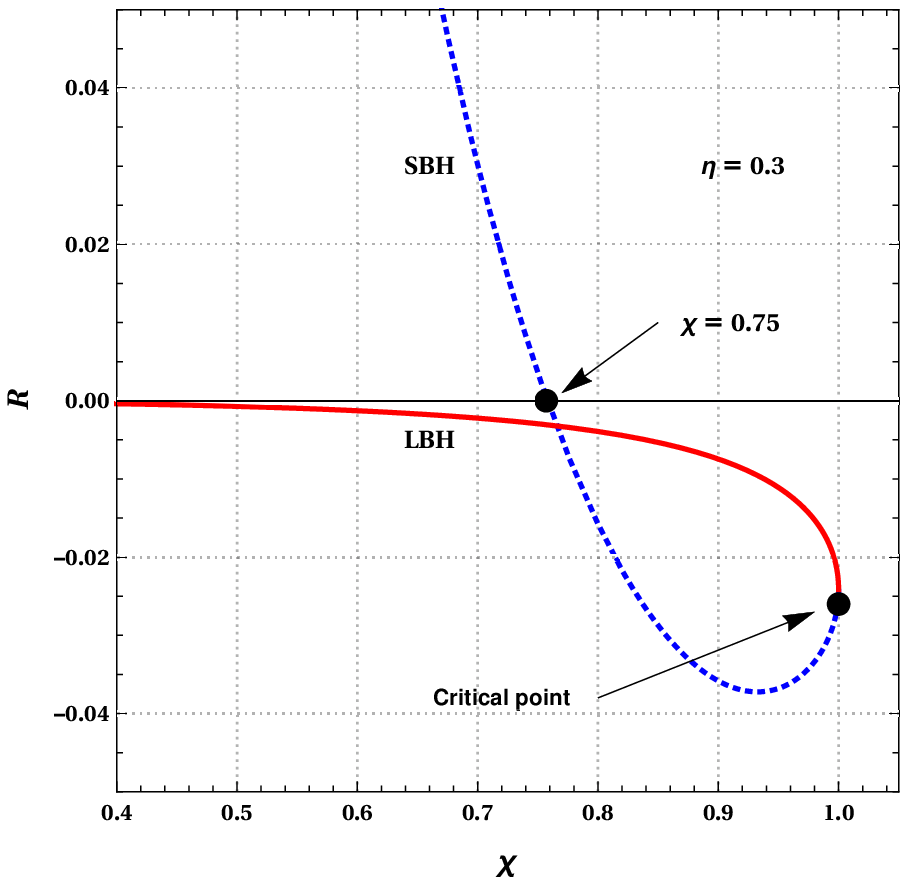}
        \label{RTE1}
    }
    \subfigure[]
    {
        \includegraphics[width=0.5\textwidth]{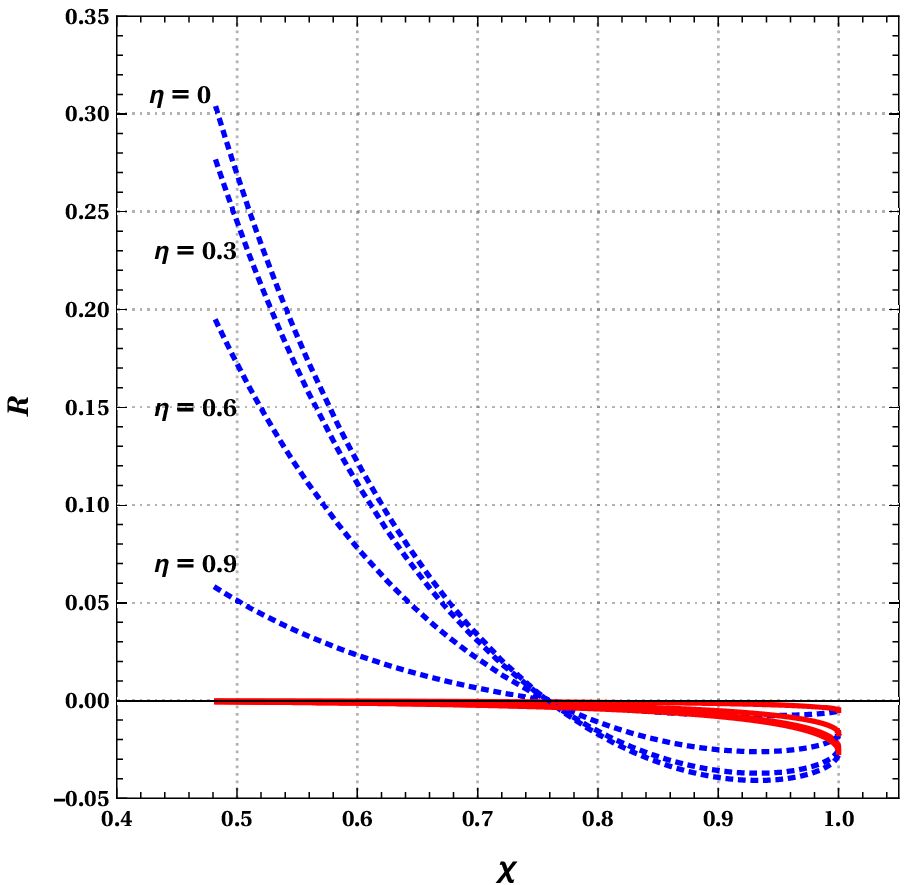}
        \label{RTE2}
    }
        \subfigure[]
    {
        \includegraphics[width=0.5\textwidth]{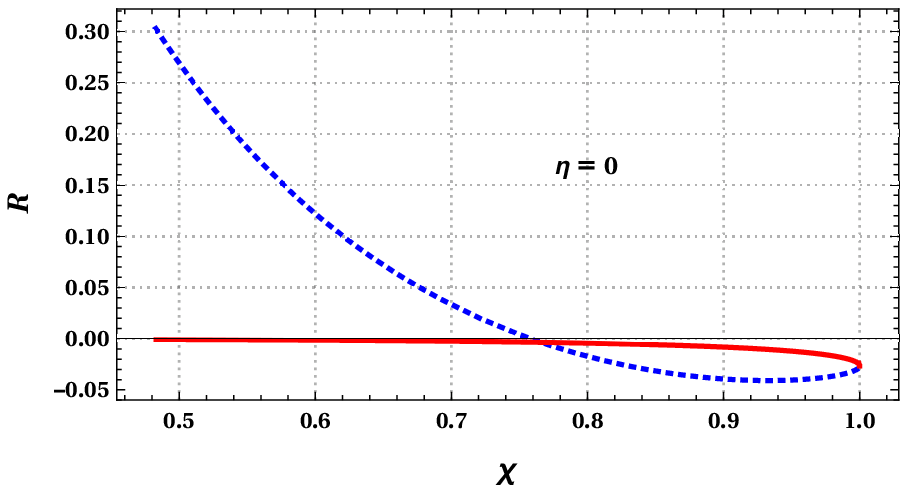}
        \label{RTE3}
    }
    \subfigure[]
    {
        \includegraphics[width=0.5\textwidth]{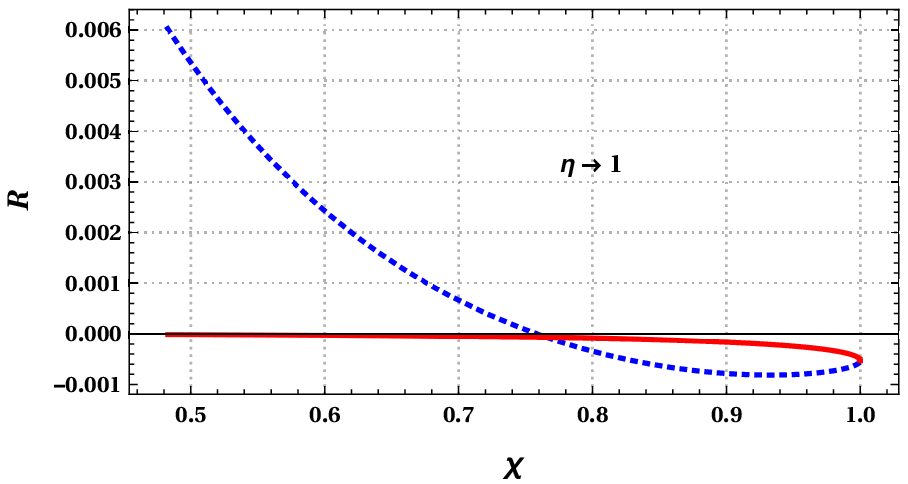}
        \label{RTE4}
    }
\\
\caption{The Ruppeiner scalar curvature $R$ along the transition curve  in extended phase space. Fig (\ref{RTE1}) is the enlarged portion near the critical point and crossing point of LBH and SBH branch. The solid red line corresponds to LBH and dotted blue line corresponds to SBH. In the second plot (\ref{RTE2}) the effect of $\eta$ is shown for different values. The last two (\ref{RTE3} and \ref{RTE4}) shows the behaviour at $\eta =0$ and $\eta \rightarrow 1$ respectively. Even though the functional behaviour remains the same when $\eta$ approaches unity the magnitude of $R$ for SBH reduces drastically. (We set $Q=1$ for each plot).  }\label{RTextended}
\end{figure}

Solving equation (\ref{Maxarea}), (\ref{addExt}) and (\ref{subExt}) simultaneously we obtain
\begin{equation}
r_2=\frac{Q}{ax}\sqrt{1+4x+x^2}
\end{equation}
\begin{equation}
P_0=\frac{3 a^2 x^2}{8 \pi  Q^2 \left(x^2+4 x+1\right)^2}
\end{equation}
and 
\begin{equation}
\chi =\frac{3 \sqrt{6} x (x+1)}{\left(x^2+4 x+1\right)^{3/2}}
\end{equation}
where $\chi =T/T_C$ $(0<\chi \leq 1)$.  Using the above results the curvature scalar for two branches can be written in terms of $x$,
\begin{equation}
R_1=-\frac{a \left(x^2+4 x-1\right)}{4 \pi  Q^2 x (x+1) \left(x^2+4 x+1\right)}
\end{equation}
\begin{equation}
R_2=\frac{a x^2 \left(x^2-4 x-1\right)}{4 \pi  Q^2 (x+1) \left(x^2+4 x+1\right)}.
\end{equation}

From this we obtain  $R-\chi$ plot which is shown in figure (\ref{RTextended}).

The LBH branch $R_2$ is always negative which indicates that the kind of interaction is like ideal Bose gas, i.e. attractive in nature. However, when the temperature decreases, the scalar curvature approaches zero, implying that the interaction strength decreases and the interaction changes from ideal Bose like gas to a classical ideal gas. Therefore, the constituents of extremal LBH is much like classical ideal gas particles. The SBH branch $R_1$ vanishes at $T=0.75 T_C$, negative for $T>0.75T_C$ and is positive for $T<0.75T_C$. From this, we can conclude that the constituents of SBH behave like anyon gas in non-extremal case. However, in the extremal case, it resembles the ideal Fermi gas. Both LBH and SBH share two points with the same $R$ value, one at the critical point and the other at the crossing point. At both points $R$ is negative and at the critical point $R_C=a/(12 \pi Q^2)$. From the $R-\chi$ plots, we can say that, for $0.75T_C<T<T_C$  the outward degenerate pressure due to repulsive nature of the interaction in SBH will expand it and form LBH, which is a first-order phase transition. On the other hand, in the region $T<0.75 T_C$  such an interpretation is not feasible because both have attractive interaction. The role of $\eta$ can be seen as a parameter which affects only the extremal SBH, without affecting the extremal LBH. In the non-extremal stretch of both SBH and LBH, $\eta$ has no significant effect.

\subsection{Alternate  Phase Space Geometry}

The $R-\chi $ plot can be used to analyse the microstructure of the black hole in alternate approach also \citep{Dehyadegari2017, Chabab2018} . The curvature scalar in alternate phase space is obtained by choosing $Y^i=\{S,Q^2\}$ as coordinates, 
\begin{equation}
R=\frac{\pi  a+6 S}{3 S^2+\pi  a S-\pi ^2 Q^2}.
\end{equation}
The  expression for two branches of $R$ for SBH and LBH are constructed from Maxwell's equal area law in alternate phase space
\begin{equation}
Q^2_0(\Psi_2-\Psi_1)=\int _{r_1}^{r_2} Q^2d\Psi .
\end{equation}
In $Q^2-\Psi$ plane, two points $(Q_0^2,\Psi _1)$ and $(Q_0^2,\Psi _2)$ in the isotherm $T=T_0$ are joined by a line under Maxwell construction. These points correspond to SBH and LBH with event horizon $r_1$ and $r_2$ at first order transition point. The above integral reduces to 
\begin{equation}
Q^2_0=a^2\left[ -2\pi T_0 xr_2^3(1+x)+xr_2^2+\frac{xr_2^4}{l^2}(1+x+x^2)\right]
\label{Maxarea2}
\end{equation}
where we have taken $r_1=xr_2$ as before. Writing equation of state (\ref{eos alternate}) at those points,
\begin{equation}
Q_0 ^2=a^2\left[ \frac{3r_1 ^4}{l^2}+r_1 ^2-4\pi r_1 T_0\right]
\end{equation}
and 
\begin{equation}
Q_0 ^2=a^2\left[ \frac{3r_2 ^4}{l^2}+r_2 ^2-4\pi r_2 T_0\right].
\end{equation}
Equating the R.H.S. of the above two equations we get
\begin{equation}
0=\left[\frac{3}{l^2}r_2^2(1+x^2)+1\right](1+x)-4\pi T_0 r_2(1+x+x^2) 
\label{sub2} 
\end{equation} 
and adding the same set of equations we have, 
\begin{equation}
2Q_0^2=a^2\left[ \frac{3}{l^2}r_2^4(1+x^4)+r_2^2(1+x^2)-4\pi T_0 r_2^3(1+x^3)\right].
\label{add2}
\end{equation}
Combining equations (\ref{Maxarea2}), (\ref{sub2}) and (\ref{add2}), with a new variable $\chi =T/T_C$ ($0<\chi <1$), we get
\begin{equation}
r_2=\frac{l}{\sqrt{1+4x+x^2}},
\end{equation}
\begin{equation}
Q^2=\frac{a^2 l^2 x^2}{\left(x^2+4 x+1\right)^2}
\end{equation}
and
\begin{equation}
\chi =\sqrt{\frac{3}{2}} \frac{ (x+1)}{\sqrt{x^2+4 x+1}}.
\end{equation}

Finally we have the two branches of curvature scalar in terms of $x$,

\begin{equation}
R_1=\frac{\left(x^2+4 x+1\right) \left(7 x^2+4 x+1\right)}{4 \pi  a x^3 (x+1)}
\end{equation}

\begin{equation}
R_2=\frac{\left(x^2+4 x+1\right) \left(x^2+4 x+7\right)}{4 \pi  a (x+1)}.
\end{equation}

\begin{figure}[H]
    \subfigure[]
    {
        \includegraphics[width=0.5\textwidth]{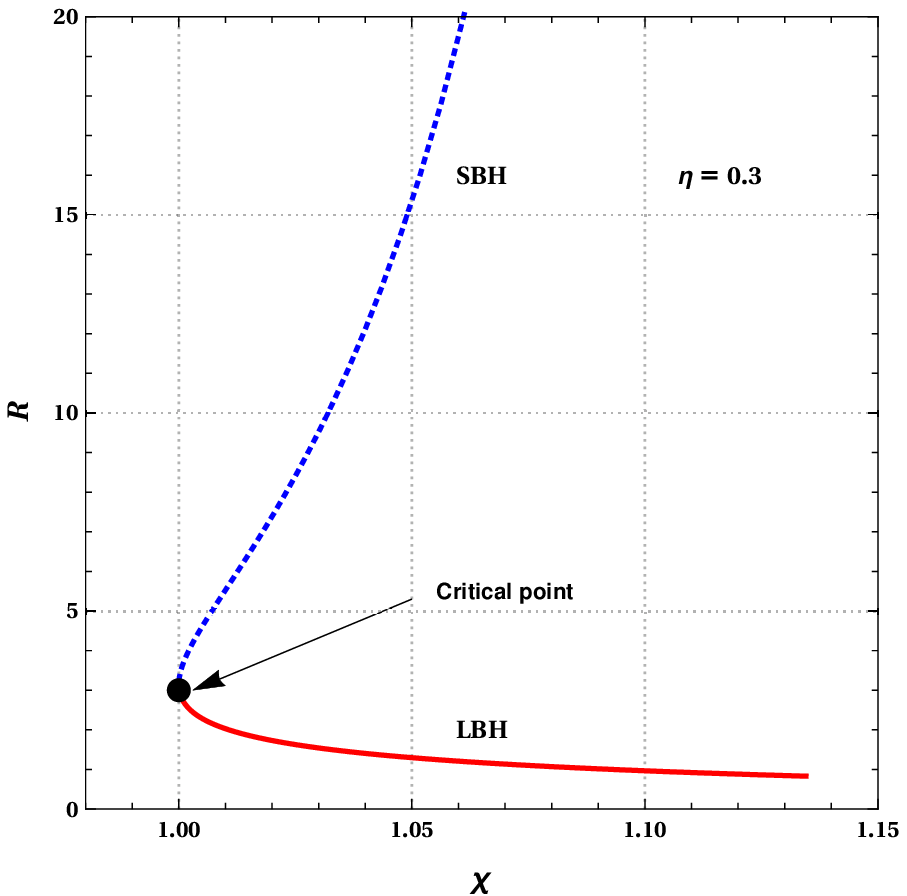}
        \label{RT1}
    }
    \subfigure[]
    {
        \includegraphics[width=0.5\textwidth]{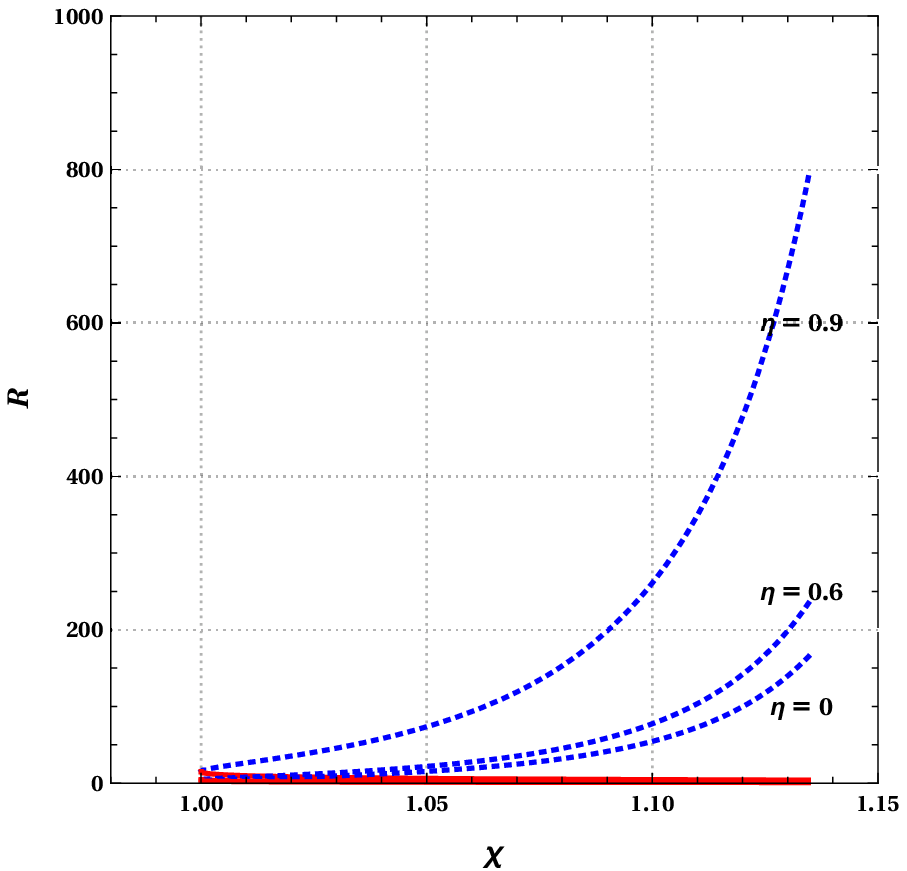}
        \label{RT2}
    }
        \subfigure[]
    {
        \includegraphics[width=0.5\textwidth]{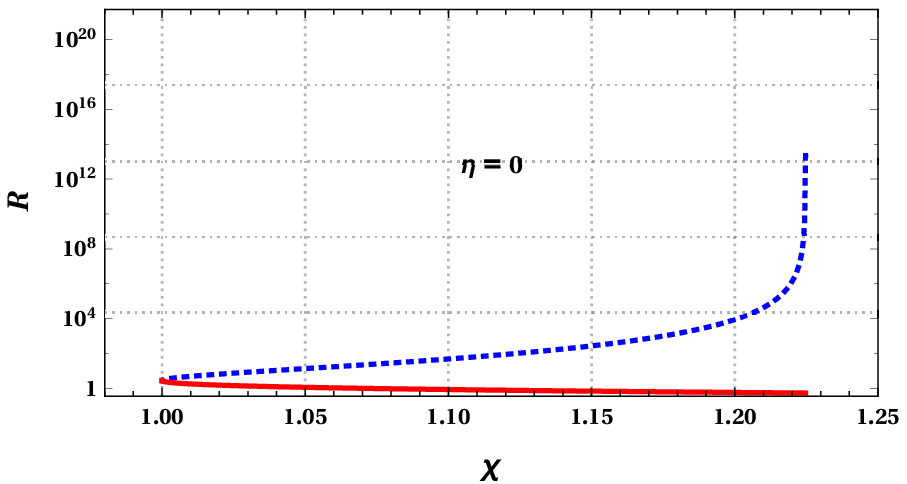}
        \label{RT3}
    }
    \subfigure[]
    {
        \includegraphics[width=0.5\textwidth]{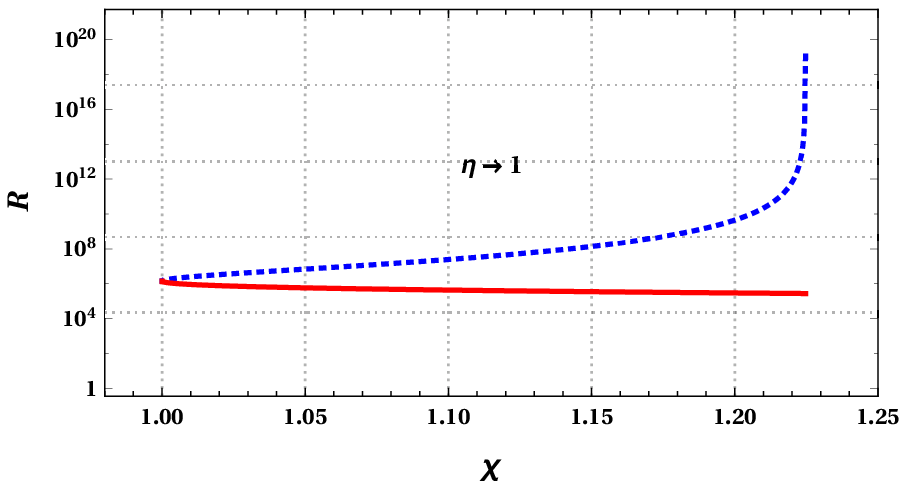}
        \label{RT4}
    }
\\
\caption{The Ruppeiner scalar curvature $R$ along the transition curve  in alternate phase space. Figure (\ref{RT1}) is the zoomed in portion near the critical point. Blue dotted line represents SBH phase and solid red line corresponds to LBH phase. In the figure (\ref{RT2}) the effect of $\eta$ is depicted for different values. Figures (\ref{RT3}) and (\ref{RT4}) are drawn in logarithmic scale. In the limit $\eta \rightarrow 1$ both branches shift equally but the characteristic behaviour remains the same.}\label{RTaternate}
\end{figure}

From this, we have $R-\chi$ plot which is shown in figure (\ref{RTaternate}).

The behaviour of $R$ in alternate space is different from that of extended phase space. Here, both $R_1$ and $R_2$ are always positive, and hence the interaction is repulsive for LBH as well as SBH. The gap between two phases widens starting from a critical point and they never cross each other. In the SBH branch, $R$ increases rapidly as the temperature reaches $T=1.22T_C$. The SBH to LBH transition can be interpreted as the continuous expansion driven by the degenerate pressure in the interior of SBH. Since very large $R$ indicates a strong repulsion among the constituents, we can conclude that the black hole near $T = 1.22T_{C}$ behaves just like an ideal Fermi gas near absolute temperature (where the degenerate pressure due to Fermi exclusion principle dominates over the thermodynamic interaction). The presence of $\eta$ enhances this behaviour to a larger extent. The increasing gap between the branches for larger values of $\eta$ is a clear indication that it is a hindrance for the phase transition. The LBH branch is less affected for all values of $\eta$ except when $\eta \rightarrow 1$. In the maximal strength of $\eta$, both LBH and SBH branches are shifted to larger values, i.e., the repulsive interaction in the interior is higher.

\section{Results and Discussion}
\label{last section}

In this article, we have carried a detailed study on the phase transition and thermodynamic geometry of charged AdS black hole with a global monopole. The isothermal study, Gibbs free energy plots, coexistence curve and the behaviour of thermodynamic curvature scalar is analysed in two different approaches, namely, in extended and alternate phase spaces. In both the spaces, we find that the black hole shows a first-order phase transition analogous to van der Waals liquid-gas system. The presence of global monopole parameter $\eta$ in the equation of state has a significant effect on the thermodynamics and the corresponding geometry. In summary, the solid angle deficit induced by $\eta$ changes the microstructure of the black hole resulting in reduced critical behaviour.

In the extended phase space, the $P-v$ diagram loses its characteristic behaviour with the increase in strength of $\eta$ (figure \ref{PVs}). This is because the van der Waals like behaviour is getting suppressed by $\eta$. Also, when $\eta$ becomes zero, the result obtained in RN-AdS black hole can be recovered.  However, the thermodynamic potential $G$ shows a swallow tail behaviour below a critical temperature, indicating that the first-order transition persists for all values of the monopole parameter. We also observe that, for high-pressure values, SBH is preferred and for the low-pressure region, LBH is favoured. The parameter $\eta$ gradually decreases the region available for LBH phase and at its maximal strength, only a negligible region is accessible for LBH (figure \ref{GPplot1}). With the help of a parametric solution from Maxwell's equal-area law, we obtained the coexistence curve. The coexistence region depends on the symmetry breaking parameter $\eta$,  which decreases with an increase of $\eta$ (figure \ref{Coexteta_extended}). The critical exponents are also calculated, which are unaffected by $\eta$ and matches with their universality class. Finally, the microscopic structure of the black hole is investigated, where the effect of the solid angle deficit induced by $\eta$ is studied. The sign of the curvature scalar is negative for LBH and approaches to zero as $T\rightarrow 0$. Whereas, for SBH it has both negative and positive sign in different regions and diverges as $T\rightarrow 0$. Making an analogy to quantum gases, the LBH interaction resembles a Bose gas and the SBH interaction resembles that of an anyon gas. However, in the extremal limit, SBH looks like an ideal Bose gas and LBH like an ideal gas. The presence of $\eta$ decreases the \emph{Bose-like} behaviour of extremal SBH and do not affect extremal LBH.

In the second approach, we have studied thermodynamics and the microstructure of the same black hole in alternate phase space. 
Here $Q^2$ is considered as a thermodynamic variable keeping cosmological constant fixed. The black hole shows similar behaviour as in extended space, confirming the existence of a first-order transition. The isotherms are studied in $Q^2-\Psi$ plane which shows the oscillatory behaviour. In contrast to extended phase space, where the unstable region is for $T<T_C$, here it is seen for $T>T_C$. However, this behaviour is comparable to van der Waals fluid. In this alternate point of view too, the system is influenced by the presence of $\eta$ which tries to diminish the critical behaviour. We calculated the critical quantities $T_C$, $P_C$ and $Q^2_C$ among which, interestingly, $T_C$ is independent of monopole parameter. The Gibbs free energy is obtained as a function $G(Q^2, T)$ and is plotted against $Q^2$. It shows swallow tail behaviour for $T>T_C$ confirming the earlier contrasting result. For higher values of $Q^2$, SBH phase is favoured whereas, for lower values, LBH phase is preferred. Increasing $\eta$ reduces the $G$ value of LBH phase by bringing down the swallow tail. The coexistence curve in $Q^2-T$ plane is also affected by the global monopole. Here, it is observed that the coexistence region decreases with an increase of $\eta$ as in extended space. We also note that the coexistence phase exists only above the critical temperature, $T_C$. In the end, we studied the microstructure in alternate phase space by analysing $R-\chi$ curve. In contrast to the extended phase space, both the LBH and SBH phases have positive curvature scalar with the same kind of interaction. The positive sign of $R$ indicates that the constituents of the black hole behave as Fermi gas with a strong repulsive force within it. At $T=1.22T_C$ curvature scalar becomes very large and hence the situation is much like Fermi gas as $T\rightarrow 0$. The parameter $\eta$ influences predominantly in SBH branch than LBH branch except for the limit $\eta \rightarrow 1$ where both branches shift together significantly. Otherwise, $\eta$ increases the gap between these two phases making the phase transition less likely.

In the extended and alternate phase space studies of the black hole, the phase structure results are similar but they differ in inferring the microstructure. This discrepancy may be due to the lack of proper formalism for understanding the black hole constituents. In Ruppeiner geometry, the curvature scalar will give an indication only about the type of interaction but not about the black hole constituents. The identifications made in this work about the nature of the black hole microstructure interaction with Fermi, Bose and anyon gases are phenomenological and the exact interpretation is still an open question. However, we assert that the presence of global monopole $\eta$ will play a significant role in influencing the microstructure of the black hole irrespective of the formalism.

\section*{Acknowledgments}
Authors Ahmed Rizwan C.L and Naveena Kumara A. acknowledge the help of Kartheek Hegde and Shreyas Punacha in preparation of the manuscript. Author N.K.A. would like to thank U.G.C. Govt. of India for financial assistance under UGC-NET-JRF scheme.

%\section*{References}


 %\bibliographystyle{elsarticle-num} 
  \bibliography{BibTex}


%\begin{thebibliography}{00}


%\bibitem{}

%\end{thebibliography}
\end{document}